\documentstyle[preprint,prb,aps]{revtex}

\begin{document}

\draft

\title{Microscopic Analysis of the Non-Dissipative Force on a
       Line Vortex in a Superconductor:
       Berry's Phase, Momentum Flows and the Magnus Force}
\author{Frank Gaitan\thanks{Address after October 1st, 1994: International
         Center for Theoretical Physics; P. O. Box 586; 34100 Trieste, ITALY.}}
\address{Department of Physics; University of British Columbia;
         Vancouver, British Columbia, CANADA; V6T 1Z1}
\date{\today}

\maketitle

\begin{abstract}
A microscopic analysis of the non-dissipative force ${\bf F}_{nd}$ acting
on a line vortex in a type-II superconductor at $T=0$ is given. We first
examine the Berry phase induced in the {\em true\/} superconducting ground
state
by movement of the vortex and show how this phase introduces a Wess-Zumino
term in the hydrodynamic action $S_{hyd}$ of the
superconducting condensate. Appropriate variation of $S_{hyd}$ gives
${\bf F}_{nd}$ and variation of the Wess-Zumino term
is seen to contribute the Magnus (lift) force of classical
hydrodynamics to
${\bf F}_{nd}$. Because our analysis is based on the {\em true\/}
superconducting ground state, we are able to confirm and strengthen earlier
work by Ao and Thouless which examined the Berry phase arising in an ansatz
for the many-body ground state. We also determine ${\bf F}_{nd}$ through a
microscopic derivation of the continuity equation for the
condensate linear momentum. This equation yields the acceleration
equation for the superflow and shows that the vortex acts as a sink for the
condensate linear momentum. The rate at which momentum is lost to the
vortex determines the non-dissipative force ${\bf F}_{nd}$ and the result
obtained agrees identically with the Berry phase calculation. The Magnus force
contribution to ${\bf F}_{nd}$ is seen in both calculations to be a consequence
of the vortex topology. A preliminary discussion is given regarding finite
temperature extensions of the Berry phase calculation, with emphasis on its
relevance for the sign anomaly occurring in Hall effect experiments on
type-II superconductors in the flux flow regime.
\end{abstract}


\section{Introduction}
As is well-known, the mixed state of a type-II superconductor is
characterized by the partial penetration of magnetic flux into the
superconductor in the form of flux lines, or
vortices. Efforts to understand the dynamical behavior of such vortices
have persisted for almost 30 years, and though much progress has been
made, a number of basic issues continue to be controversial. A vortex
in a type-II superconductor is acted on by 3 classes of forces:
(i) pinning forces due to lattice defects; (ii) dissipative forces
due to coupling of the vortex core to the lattice; and (iii) a
non-dissipative force ${\bf F}_{nd}$ due to an applied magnetic field
${\bf H}_{ext}$, an electric field ${\bf E}$ generated by the vortex motion,
and the hydrodynamic pressure of the surrounding condensate of
superconducting electrons. Already in the classic models of
Bardeen-Stephen (BS) \cite{bs} and Nozi\`{e}res-Vinen (NV) \cite{nv},
the form of ${\bf F}_{nd}$ is controversial. The debate centers
around whether ${\bf F}_{nd}$ includes a contribution from the
Magnus (lift) force of classical hydrodynamics which acts on a solid
body moving through a fluid which circulates about it \cite{bat}.
Both models: (1) are macroscopic/phenomenological in character, based
on Maxwell-London electrodynamics, classical thermodynamics, and
physical intuition; (2) assume strongly type-II superconductors so
that the non-local character of BCS superconductivity can be approximated
by a local dynamics; (3) assume $T=0$ so that normal electrons are
only present inside the vortex core whose radius is equal to the
zero temperature coherence length $\xi_{0}$; and (4) the applied
magnetic field satisfies $H_{c_{1}}<H_{ext}\ll H_{c_{2}}$ so that
vortex-vortex interactions can be ignored. The vortex is assumed to
be immersed in an applied transport current
${\bf J}=\rho_{s}e{\bf v}_{s}$, where $\rho_{s}$ is the superconducting
electron density far from the vortex; $e$ is the electron charge;
and ${\bf v}_{s}$ is the velocity of the applied supercurrent with
respect to the lattice rest frame. In the BS model, the non-dissipative
force (per unit length) is due to the Lorentz force,
$\rho_{s}h\omega ({\bf v}_{s}\times\hat{{\bf z}})/2$. Here $h$ is
Planck's constant; and $\omega =\pm 1$ is the vortex winding number
whose sign specifies the sense of the condensate superflow
about the vortex (which is threaded by a single flux quantum $\phi_{0}
=hc/2e$). In this paper it will be assumed that all vortices are
rectilinear (viz.\ line) vortices whose axis lies along
$\hat{{\bf z}}$. In the NV model, ${\bf F}_{nd}$ is the sum of the
Lorentz force and the Magnus (lift) force $-\rho_{s}m{\cal K}
{\bf v}_{L}\times\hat{{\bf z}}$, where $m$ is the electron mass;
${\cal K}=h\omega/2m$ is the circulation of the condensate near the
vortex core; and ${\bf v}_{L}$ is the translational velocity of the
line vortex with respect to the lattice. The discrepancy in the form
of ${\bf F}_{nd}$ in the two models leads to different predictions for
the Hall angle in the flux flow regime \cite{ks}. Recent work by
Ao and Thouless \cite{ath} (A\&Th) has renewed interest in this
controversy concerning ${\bf F}_{nd}$. These authors argue that the
correct form for ${\bf F}_{nd}$ is the NV-form, ${\bf F}_{nd}=
\rho_{s}h\omega ({\bf v}_{s} - {\bf v}_{L})\times \hat{{\bf z}}/2$,
and that the Magnus (lift) force contribution arises from a Berry
phase induced in the many-body ground state by the vortex motion.
Their analysis is based on an ansatz for the many-body ground state and
so it might be objected that their result concerning the presence of
the Magnus force is a consequence of their ansatz and not an actual
property of the true superconducting ground state. Thus it is of
considerable interest to see if ${\bf F}_{nd}$ can be determined on the
basis of a microscopic analysis in which the superconducting
dynamics is treated exactly.

In this paper, two calculations of the non-dissipative force
${\bf F}_{nd}$ will be provided which are based on the microscopic
formulation of the superconducting dynamics due to Bogoliubov
\cite{deg}. This approach is powerful enough to treat problems with spatial
inhomogeneities as occur when a vortex is present in the
superconductor. In Section~\ref{sec2}, ${\bf F}_{nd}$ will be
determined by working with the true superconducting ground state in
the case where a vortex is present. We first show how this state is constructed
from the {\em exact\/} solutions of the Bogoliubov equations in the
presence of a vortex. This state is seen to develop a Berry phase
as a consequence of the vortex motion. Our approach in this subsection
is a significant generalization of one used previously to determine
the intrinsic orbital angular momentum in $^{3}{\rm He}$--A for
a spatially uniform orbital texture \cite{us}. The Berry phase is
calculated and found to agree with the result of Ref.~\onlinecite{ath}
in the case of a neutral superfluid.
A discussion of why the approximate calculation of A\&Th has produced the
exact result is also given. In the case of a charged superfluid,
gauge invariance requires a correction to the Berry phase obtained in
Ref.~\onlinecite{ath}; however this correction will be seen to not influence
the final result for ${\bf F}_{nd}$.
It is then shown how the Berry phase induces a Wess-Zumino term in the
action describing the hydrodynamic degrees of freedom of the
superconducting condensate.
Variation of the hydrodynamic action with respect to the vortex
trajectory gives ${\bf F}_{nd}$ which is found to have the NV-form.
This calculation also shows that the Magnus force contribution to
${\bf F}_{nd}$ {\em is\/} a consequence of the Berry phase induced
by the vortex motion. As our calculation is based on the {\em true}
superconducting ground state, we see that the result of A\&Th is
{\em not} a consequence of their ansatz, but a real property of
type-II superconductivity. In Section~\ref{sec3}, ${\bf F}_{nd}$ is
determined via a microscopic derivation of the acceleration
equation for the superflow. We find the expected contributions off
the vortex due to spatial variation of the chemical potential, and to
the electric and magnetic fields present. But by carefully tracking
the effects of the vortex topology on the dynamics, we also find
a singular term that describes the disappearance of linear momentum
into the vortex, and which represents the non-dissipative force
acting on the vortex. We find that this force is also given by the
NV-form of ${\bf F}_{nd}$, in agreement with the Berry phase
calculation of Section~\ref{sec2}. As NV have shown \cite{nv}, the
terms in the acceleration equation acting off the vortex lead to a
flux of linear momentum in towards the vortex and corresponds to
their determination of ${\bf F}_{nd}$. By keeping track of topological
effects, we will see that the rate at which linear momentum is appearing
inside the vortex is exactly equal to the rate at which it is flowing
in towards it so that linear momentum conservation is maintained in the
combined condensate-vortex system as expected.
We close in Section~\ref{sec4} with a discussion of our
results; and with preliminary remarks concerning finite temperature effects
and their possible consequences for the sign anomaly occurring in
Hall effect experiments on type-II superconductors in the
flux flow regime. Two Appendices are also included: Appendix A indicates how
the Berry phase appears in a propogator formulation of the Adiabatic Theorem;
while Appendix B gives a microscopic derivation of the gauge invariant
Wess-Zumino term which is independent of the Berry phase calculation given in
Section \ref{sec2}, and also obtains the hydrodynamic action for the
superconducting condensate.

\section{Berry's Phase and the Non-Dissipative Force ${\bf F}_{nd}$}
\label{sec2}
This Section is organized as follows. In Section~\ref{sec2a} we construct
the true superconducting ground state in the case where a single vortex is
present and derive an expression for the Berry phase induced in this state
by the vortex motion in terms of the Berry phase induced in the
eigenstates of the Bogoliubov equation (in the presence of a
single vortex). In Section~\ref{sec2b} the Berry phase in the true
superconducting ground state is evaluated. In Section~\ref{sec2c} we show
how this Berry phase induces a topological Wess-Zumino term in
the action describing the hydrodynamic degrees of freedom of the
superconducting condensate. Variation of the hydrodynamic action with respect
to the vortex trajectory gives ${\bf F}_{nd}$ and it is seen in the course of
evaluating this variation that the Berry phase induced Wess-Zumino term
is responsible for the occurence of the classical Magnus (lift) force in
${\bf F}_{nd}$.

\subsection{Constructing the Ground State}
\label{sec2a}
Our starting point is the Bogoliubov equation \cite{deg}. This formulation
of the quasiparticle dynamics is capable of handling the spatial
inhomogeneities introduced by a line vortex which for the time being is
assumed to be fixed at the origin (in the lattice rest frame). As in the models
of BS and NV, we will: (i) assume $T=0$ so that the physical behavior of the
superconductor is captured by the superconducting ground state; (ii)
approximate the non-local character of the BCS superconducting ground state
by a local dynamics which is, strictly speaking, only true for strongly
type-II superconductors (it is not anticipated that non-local effects will
qualitatively modify our results); and (iii) assume
$H_{c_{1}}<H_{ext}\ll H_{c_{2}}$
so that vortex-vortex interactions can be ignored and we can focus on a
single vortex. We adopt cylindrical coordinates ($r$, $\theta$, $z$) with
the z-axis along the vortex axis. In the presence of a line vortex, the
gap function takes the form $\Delta({\bf r})=\Delta_{0}(r)\exp[-i\theta]$ so
that the phase of the gap $\phi({\bf r})=-\theta({\bf r})$ is not a
single valued function of position. As the field point ${\bf r}$ encircles
the vortex axis, $\phi\rightarrow\phi - 2\pi$ (more generally, $\Delta({\bf r})
=\Delta_{0}(r)\exp[i\omega\theta]$ so that $\phi\rightarrow\phi + 2\pi\omega$,
where $\omega$ is an integer known as the vortex winding number). The gap
amplitude $\Delta_{0}(r)$ goes linearly to zero at the origin and approaches
a constant value far from the vortex. In the presence of a magnetic field
${\bf H}=\nabla\times{\bf A}$, the Bogoliubov equation is
\begin{equation}
\left( E_{n} - H_{BOG}\right) \left(\begin{array}{c}
                                       u_{n} \\
                                       v_{n}
                                    \end{array}\right) = 0 \hspace{0.1in} ,
\label{bog}
\end{equation}
where
\begin{equation}
H_{BOG} =                       \left(\begin{array}{cc}
                                         \frac{1}{2}\left(-i\nabla -
                                          e{\bf A}\right)^{2}
                                           + U_{0}({\bf r}) - E_{f} &
                                            \Delta({\bf r}) \\
                                         \Delta^{\ast}({\bf r}) &
                                          -\frac{1}{2}\left(i\nabla -
                                           e{\bf A}\right)^{2}-
                                            U_{0}({\bf r}) +E_{f}
                                      \end{array}\right)
\hspace{0.1in} .
\end{equation}
Here $E_{f}= p_{f}^{2}/2$ is the Fermi energy ($\hbar =m=c=1$ unless
explicitly stated otherwise); $n$ is an index
which labels the spectrum of $H_{BOG}$; $U_{0}({\bf r})$ is the pinning
potential which will henceforth be set equal to zero so that we are dealing
with a clean type-II superconductor. As is well-known, equation~(\ref{bog})
has positive and negative energy solutions (relative to the Fermi surface)
which are related as follows: if ($u_{n}$ $v_{n}$) is (the adjoint of) a
solution of equation~(\ref{bog}) with energy $E_{n}>0$, then ($-v^{\ast}_{n}$
$u^{\ast}_{n}$) is (the adjoint of) a solution of equation~(\ref{bog}) with
energy $-E_{n}<0$. The two-component elementary excitations built up from
the solutions of equation~(\ref{bog}) will be referred to as Nambu
quasiparticles
(NQP) and quasi-holes (NQH). Let $a^{\dagger}_{n, 1}$ create a positive
energy ($E_{n}>0$) NQP and $a^{\dagger}_{n,2}$ create a negative energy
($E_{n}<0$) NQP. Introducing the two-component field operator $\Psi({\bf r})$
for a NQP, we expand it in terms of the solutions of eqn.~(\ref{bog})
\begin{equation}
\Psi({\bf r})= \sum_{n}\left[\: a_{n,1}\left(\begin{array}{c}
                                                u_{n} \\
                                                v_{n}
                                             \end{array}\right) +
                                a_{n,2}\left(\begin{array}{c}
                                                -v^{\ast}_{n} \\
              \label{tran2}                     u^{\ast}_{n}
                                             \end{array}\right)\:\right]
\hspace{0.1in} .
\end{equation}
$\Psi({\bf r})$ can also be defined in terms of the field operators for the
spin-up and spin-down Landau quasiparticles (LQP)
\begin{equation}
\Psi({\bf r})\equiv \left(\begin{array}{c}
                             \psi_{\uparrow}({\bf r}) \\
                             \psi_{\downarrow}^{\dagger}({\bf r})
                          \end{array}\right) =
                        \sum_{n}\left[\: \gamma_{n\uparrow}
                         \left(\begin{array}{c}
                                  u_{n} \\
                                  v_{n}
                               \end{array}\right) +
                         \gamma^{\dagger}_{n\downarrow}
                          \left(\begin{array}{c}
                                   -v^{\ast}_{n} \\
               \label{tran}        u^{\ast}_{n}
                                \end{array}\right)\:\right] \hspace{0.1in} .
\end{equation}
The right-most part of eqn.~(\ref{tran}) comes from expanding the LQP field
operator in terms of the creation and annihalation operators
$\{\: \gamma_{n\uparrow},\; \gamma^{\dagger}_{n\uparrow};\;
\gamma_{n\downarrow}, \; \gamma^{\dagger}_{n\downarrow}\:\}$ for the
Bogoliubov quasiparticles (BQP) \cite{deg}. Comparison of eqns.~(\ref{tran2})
and (\ref{tran}) indicates that $a_{n,1}=\gamma_{n\uparrow}$ and
$a_{n,2}=\gamma^{\dagger}_{n\downarrow}$. Thus the spin projection
$2s_{z}$ labels the charge conjugation degree of freedom of the NQP's
\cite{nam} and the superconducting ground state is obtained by occupying the
negative energy states
\begin{equation}
|BCS\rangle = \prod_{n}\: a^{\dagger}_{n,2}|0\rangle =
 \label{gsf}    \prod_{n}\: \gamma_{n\downarrow}|0\rangle \hspace{0.1in} ,
\end{equation}
where $|0\rangle$ is the zero-particle state.

As $H_{BOG}$ is Hermitian, the solutions of eqn.~(\ref{bog}) satisfy
orthogonality and completeness relations \cite{bar}. The orthogonality
relations are:
\begin{eqnarray}
\int\: d^{3}r\:\left[\: u_{n}({\bf r})u^{\ast}_{m}({\bf r})+v_{n}({\bf r})
        v^{\ast}_{m}({\bf r})\: \right] & = & \delta_{nm} \nonumber \\
\int\: d^{3}\left[ \: u_{n}({\bf r}) v_{m}({\bf r}) - v_{n}({\bf r})
        u_{m}({\bf r})\: \right] & = & 0 \hspace{0.1in} ;
\end{eqnarray}
and the completeness relations are:
\begin{eqnarray}
\sum_{n}\: \left[\: u_{n}({\bf r})u^{\ast}_{n}({\bf r}^{\prime}) +
      v^{\ast}_{n}({\bf r})v_{n}({\bf r}^{\prime})\: \right] & = &
       \delta({\bf r}- {\bf r}^{\prime}) \nonumber \\
\sum_{n}\: \left[\:u_{n}({\bf r})v^{\ast}_{n}({\bf r}^{\prime}) -
      v^{\ast}_{n}({\bf r})u_{n}({\bf r}^{\prime}) \: \right] & = &
       0 \hspace{0.1in} .
\end{eqnarray}
Use of the orthogonality relations allows us to invert eqn.~(\ref{tran})
yielding
\begin{eqnarray}
\gamma_{n\downarrow} & = & \int\: d^{3}r\: \left[ \: -\psi^{\dagger}_{\uparrow}
    ({\bf r})v^{\ast}_{n}({\bf r}) + \psi_{\downarrow}({\bf r})
     u^{\ast}({\bf r})\: \right] \nonumber \\
\gamma^{\dagger}_{n\uparrow} & = & \int\: d^{3}r\: \left[ \:
    \psi^{\dagger}_{\uparrow}({\bf r})u_{n}({\bf r}) +
     \psi_{\downarrow}({\bf r})v_{n}({\bf r})\: \right]
   \hspace{0.1in} . \label{orth}
\end{eqnarray}

As we shall see in the following sub-section, adiabatic motion of the vortex
produces a Berry phase \cite{ber} in the solutions of eqn.~(\ref{bog}),
$(\, u_{n}\; v_{n}\, )\rightarrow \exp [i\phi_{n} ](\, u_{n}\; v_{n}\, )$,
where $\phi_{n}$ depends only on the vortex trajectory ${\bf r}_{0}(t)$.
{}From eqn.~(\ref{orth}) we see that $\gamma_{n\downarrow}$ inherits this
phase,
$\gamma_{n\downarrow}\rightarrow\exp [-i\phi_{n}]\gamma_{n\downarrow}$.
Consequently (see eqn.~(\ref{gsf})), the superconducting ground state develops
a Berry phase $\Gamma$, $|BCS\rangle \rightarrow \exp [i\Gamma ]|BCS\rangle $,
where
\begin{equation}
\Gamma = -\sum_{n}\:\phi_{n} \hspace{0.1in} \label{bigber}.
\end{equation}
The physical significance of $\Gamma$ will be discussed below. First we
evaluate $\{ \phi_{n} \}$.

\subsection{Calculating the Berry Phase}
\label{sec2b}
Berry \cite{ber} showed that when a quantum system is coupled to an
adiabatically-evolving environment and the system is initially prepared in an
eigenstate of the $t=0$ Hamiltonian $H[\, {\bf R}(0)]$ (the adiabatic coupling
of the system to its environment enters through the set of parameters
${\bf R}(t)=(R_{1}(t),\cdots , R_{n}(t))$ which appear explicitly in the
system Hamiltonian $H[\, {\bf R}(t)]$), that the time development of this state
acquires a non-integrable phase $\gamma_{B}(t)$ now known as the Berry phase
\begin{equation}
|\psi(t)\rangle = \exp \left[ \: i\gamma_{B}(t)- i\int_{0}^{t}\:
  du\, E(u)\:\right]\,|E(t)\,\rangle \hspace{0.1in} .
\end{equation}
Here $|E(t)\,\rangle $ is the eigenstate of the instantaneous Hamiltonian
$H[\, {\bf R}(t)]$ which evolves continuously from the initial state
$|E(0)\,\rangle $, and the gauge invariant Berry phase appropriate for an
electrically charged particle is given by \cite{ahar}
\begin{equation}
\gamma_{B}(t)=\int^{t}_{0}\: d\tau\,\langle\, E|\, i \frac{d}{d \tau}
 -\frac{e}{\hbar}A_{0}(\tau)
  |E\,\rangle =  \int_{0}^{t}\: d\tau\langle\, E|
   \, i\frac{d {\bf R}}{d\tau}\cdot\nabla_{{\bf R}}-\frac{e}{\hbar}A_{0}
    |E\,\rangle \hspace{0.1in} .
\end{equation}
In Appendix A we show how the Berry phase arises in a propogator
approach to the Adiabatic Theorem which will be useful for our discussion
in Section~\ref{sec2c}. In our case, the quantum system is the two-component
NQP field $\Psi$; the environment is the superconducting condensate
described by the gap function $\Delta({\bf r})$; and the adiabatically-varying
parameters correspond to the vortex position ${\bf r}_{0}(t)=(\,x_{0}(t), \;
y_{0}(t)\,)$. To determine $\Gamma$ in eqn.~(\ref{bigber}) we must obtain the
eigenstates of $H_{BOG}$ for a given vortex position. These states were first
considered by Caroli et.\ al.\ \cite{car}, and a detailed solution for the
bound and scattering states was given by Bardeen et.\ al.\ \cite{bar}. These
eigenstates take the form
\begin{equation}
\hat{\chi}_{n}({\bf r})= \left( \begin{array}{c}
                                   u_{n} \\
\label{state}                      v_{n}
                                \end{array} \right) =
      e^{ik_{z}z}e^{i\mu\theta}e^{-i\frac{\theta}{2}\sigma_{z}}\hat{f}_{n}
        (r) \hspace{0.1in} .
\end{equation}
Here $\hbar k_{z}$ is the z-component of the quasiparticle momentum; $2\mu$
is required to be an odd-integer to insure single-valuedness of the
eigenstate $\hat{\chi}_{n}$; $\tan\theta ({\bf r}) =
(y - y_{0})/(x-x_{0})$;
and $r=\sqrt{(x-x_{0})^{2}+(y-y_{0})^{2}}$. The spinor $\hat{f}_{n}(r)$
has been examined in Ref.~\onlinecite{bar} analytically for a simple model of
the gap function, and numerically via a variational method. We will not require
its detailed form below. To simplify the notation, we have collected the
quantum numbers specifying the eigenstate into the single label $n$. Finally,
our calculation is done per unit length in the z-direction so that
$d^{3}x \rightarrow d^{2}x$. Thus
\begin{eqnarray}
\phi_{n} & = & i \int\: d^{2}x\, \int\: d\tau\, \dot{{\bf r}}_{0}\cdot
                \hat{\chi}^{\dagger}_{n}\,\nabla_{{\bf r}_{0}}
                 \hat{\chi}_{n} -\int\: d^{2}x\, \int\: d\tau\, \frac{e}{\hbar}
                  \, A_{0}(\tau)\,\chi^{\dagger}_{n}\chi_{n} \nonumber \\
 & = & \int\: d^{2}x\,d\tau\, \dot{{\bf r}}_{0}\cdot\left[ \left\{ \,
        \hat{\chi}^{\dagger}_{n}\left(-\mu + \frac{1}{2}\sigma_{z}\right)\,
         \hat{\chi}_{n}\,\right\}\,\nabla_{{\bf r}_{0}}\theta +
          \hat{f}^{\dagger}(r)\nabla_{{\bf r}_{0}}\hat{f}_{n}(r)\, \right]
           \nonumber \\
 &   & {} \hspace{0.3in}   -\int\: d^{2}x\,d\tau\, \frac{e}{\hbar}\,
           A_{0}(\tau)\,\chi^{\dagger}_{n} \chi_{n} \nonumber \\ \label{lilber}
 & \equiv & \phi^{\theta}_{n} + \phi^{r}_{n} -\int\: d^{2}x\, d\tau\,
             \frac{e}{\hbar}\, A_{0}(\tau)\,\chi^{\dagger}_{n}\chi_{n}
              \hspace{0.1in} .
\end{eqnarray}
Combining eqns.~(\ref{bigber}) and (\ref{lilber}) gives
\begin{eqnarray}
\Gamma & = &  -\sum_{n}\: \left( \phi_{n}^{\theta} + \phi_{n}^{r} \right)
               +\int\: d^{2}x\, d\tau\, \frac{e}{\hbar}\, A_{0}\,\sum_{n}\,
                \chi^{\dagger}_{n}\chi_{n} \nonumber \\
 & = & -\sum_{n}\: \left( \phi_{n}^{\theta} + \phi_{n}^{r} \right)
                +\int\: d^{2}x\, d\tau\, \frac{e}{\hbar}\, A_{0}\,\rho_{s}
           \nonumber \\
  & = & \Gamma^{\theta} + \Gamma^{r} + \int\: d^{2}x\, d\tau\, \frac{e}{\hbar}
         \, A_{0}\,\rho_{s} \hspace{0.1in} .
\end{eqnarray}
At this point it is necessary to recall a basic property of BCS
superconductivity. Specifically, that the global symmetry corresponding to
changing the phase of the LQP field operator is spontaneously broken.
Because of this, the ground state becomes infinitely
degenerate and is parameterized by a phase angle $f$
(see ref.~\onlinecite{schr}). The phase
$f$ and particle number $N$ become canonically conjugate observables and two
representations are possible for the many-body states.
In one representation, particle number is well-defined, but $f$ is not; while
in the second, $f$ is well-defined, but particle number is not \cite{pwa}. Our
calculation corresponds to the latter case where the many-body states
do not contain a definite number of particles. Because of this, a calculation
of ground state properties will require what amounts to a normal ordering
prescription. An example of the kind of normal ordering we have in mind is the
fixing of the Fermi energy through the requirement that the expectation value
of
the particle number operator in the ground state equal the number of electrons
present: $\sum_{k}\,2v^{2}_{k} = \bar{N}$. Because our calculation of $\Gamma$
is based on a ground state in which particle number is not well-defined, a
similar kind of normal ordering is necessary and can be implemented
straightforwardly.

We begin by re-writing $\Gamma^{r}$ as
\begin{equation}
\Gamma^{r} = -\int\, d\tau\,d^{2}x\, \dot{{\bf r}}_{0}\cdot\left(
 \nabla_{{\bf r}_{0}}r\right)
\label{gamr}  F(r) \hspace{0.1in} ,
\end{equation}
where $F(r)\equiv \sum_{n}\, \hat{f}^{\dagger}_{n}\frac{d}{dr}\hat{f}_{n}$.
$F(r)$ receives a divergent contribution from the scattering states present
in the Fermi sea. This divergence is regulated by introducing a temporary
cutoff in the sum over $n$ which is removed at the end of the calculation.
We now show that the finiteness of $F_{reg}(r)$ implies that $\Gamma^{r}=0$.
Writing $\dot{{\bf r}}_{0}=\nabla_{{\bf r}}\times {\bf C}$, where
${\bf C}=(\dot{x}_{0}y-\dot{y}_{0}x)\hat{{\bf z}}$, in eqn.~(\ref{gamr}) and
using standard vector identities gives
\begin{eqnarray}
\Gamma^{r}_{reg} & = & \int\, d\tau\,\left[\, \int_{S_{\infty}}\, dl\,
                        F_{reg}(r)
                        |{\bf C}|\,\hat{{\bf n}}\cdot\left( \hat{{\bf z}}
                         \times\hat{{\bf e}}_{r}\right) -
                       \int\, d^{2}x\, {\bf C}\cdot\left(\, \nabla_{r}F_{reg}
                        \times\nabla_{r}r + F_{reg}\nabla_{r}\times\nabla_{r}r
                         \right)\,\right] \nonumber \\
 & = & 0 \hspace{0.1in} ,
\end{eqnarray}
($S_{\infty}$ is the circle in the x-y plane at infinity) since: (i)
$\hat{{\bf n}}= \hat{{\bf e}}_{r}$; (ii) $\nabla_{r}F_{reg}(r)$ and
$\nabla_{r} r$ are parallel; and (iii) $r$ is a single-valued function of
$(x,\; y)$. Removing the cutoff gives $\Gamma^{r}=0$.

$\Gamma^{\theta}$ can be re-written as
\begin{eqnarray}
\Gamma^{\theta} & = & -\int\, d\tau\,d^{2}x\,\left(\,\dot{{\bf r}}_{0}\cdot
                       \nabla_{{\bf r}_{0}}\theta\,\right)\,\sum_{n}\,
          \left[\,|u_{n}|^{2}\left(-\mu + \frac{1}{2}\right)
        + |v_{n}|^{2}\left(-\mu - \frac{1}{2}\right)\,\right] \label{gamt1}\\
 & \equiv & - \int\, d\tau\, d^{2}x\, \left(\, \dot{{\bf r}}_{0}\cdot
 \nabla_{{\bf r}_{0}}\theta\,\right)\, S \hspace{0.1in} . \label{gamt2}
\end{eqnarray}

{}From eqn.~(\ref{gamt1}) and (\ref{state}), we see that $S$ is minus the
z-component of the orbital angular momentum density (in units of
$\hbar$) present in the Fermi sea. Although our expression for $S$
is divergent because this sea is infinite, knowledge of its
physical significance makes obvious what the physically relevant
normal ordering choice is:
\begin{equation}
S_{reg}=\langle\, -{\cal L}_{z}/\hbar\, \rangle = -\hat{{\bf z}}
 \cdot {\bf r}\times (\rho_{s}m{\bf v})/\hbar \hspace{0.1in} .
\label{normod}
\end{equation}
Here ${\bf v}= \hat{{\bf e}}_{\theta}({\cal K}/2\pi r) f(r)$ is the
superflow about the vortex; ${\cal K}= h\omega /2m$ is the circulation of
the condensate near the vortex core ($\omega$ is the vortex winding number,
see above); and
\begin{equation}
f(r) = \left\{ \begin{array}{cc}
                  1 & \xi_{0}\ll r \ll \lambda \\
                  \sim r e^{-r/\lambda} & r\gg \lambda
               \end{array}\right. \hspace{0.1in} .
\end{equation}
It is extremely important to realize that this form of $f(r)$ insures that only
one flux quantum threads the vortex \cite{f+h}. Thus our normal ordering
choice, eqn.~(\ref{normod}), is fixed by the requirement that the flux
through the vortex $\Phi$ agree with our assumption $\Phi = hc/2e$. (In the
following subsection the ground state Berry phase will be related to the
occurrence of a Wess-Zumino term in the hydrodynamic action. For an
{\em independent\/} derivation of this Wess-Zumino term which does not rely on
the above normal ordering prescription, see Appendix B.)
With this choice of normal ordering
\begin{equation}
S_{reg} = -\rho_{s}\omega /2 \hspace{0.1in} ,
\end{equation}
and
\begin{equation}
\Gamma = \int\, d\tau\, d^{2}x\, \rho_{s}\left(\,
         \frac{1}{2}\dot{{\bf r}}_{0}\cdot\nabla_{{\bf r}_{0}}\theta
          - \frac{e}{\hbar}\, A_{0}\,\right)
          \hspace{0.1in} ,
\label{bigres}
\end{equation}
for $\omega = -1$, as assumed in our calculation.
Thus we see that the {\em true superconducting ground state\/} does
in fact develop a Berry phase as a consequence of the vortex motion.
We are able to obtain this Berry phase and we see that the approximate
calculation of Ref.~\onlinecite{ath} has produced the exact result in the
case of a neutral superfluid. For a charged superfluid, the gauge
invariant form of the Berry phase must be used which adds the ``$e
A_{0}$''-term to the ground state Berry phase. It is clear from our calculation
that, in the case of a neutral superfluid, the Berry phases are ultimately due
to the azimuthal symmetry of the vortex which requires the eigenstates of the
Bogoliubov equation to be eigenfunctions of the z-component of orbital
angular momentum. Examination of the ansatz of A\&Th shows that it is such
an eigenfunction and so has the essential $\theta$-dependence necessary for
a proper determination of the Berry phase in the case of a neutral superfluid.
Because we have worked with the {\em true\/} superconducting ground state
it is clear that the gauge invariant Berry phase and its physical consequences,
to be discussed below, are true properties of the superconducting ground state
when a moving vortex is present, and not the result of a specific ansatz
wavefunction.

\subsection{Berry's Phase, the Hydrodynamic Action and ${\bf F}_{nd}$}
\label{sec2c}
Our starting point is the vacuum-to-vacuum transition amplitude for a
system of electrons with an effective attractive interaction due to phonons
which is responsible for the pairing instability which characterizes
BCS superconductivity,
\begin{equation}
W=\langle\,vac;t=T|\psi(t=T)\,\rangle=\langle\,vac;t=T|U(T,0)
    |vac;t=0\,\rangle\hspace{0.1in} .
\end{equation}
Here $U(T,0)={\cal T}(\, \exp [-i\int_{0}^{T}\,d\tau H_{BCS}]\,)$; ${\cal T}$
is the time ordering operator; and the BCS Hamiltonian is
\begin{eqnarray}
H_{BCS} & = & \int\, d^{3}x\,\psi^{\dagger}_{\sigma}(x)\left[ \,-\frac{1}
               {2}\left(\nabla -ie{\bf A}\right)^{2} -E_{f} + e A_{0}\,\right]
                \psi_{\sigma}(x) \nonumber \\
        &   & {} \hspace{0.25in} -\frac{g}{2}\int\, d^{3}x\,
                 \psi^{\dagger}_{\sigma}(x)\psi^{\dagger}_{-\sigma}(x)
                  \psi_{-\sigma}(x)\psi_{\sigma}(x) \nonumber \\
  & & \mbox{}\hspace{0.5in} +\int\, d^{3}x\,\frac{1}{8\pi}\left[ \,
       \left({\bf H}- {\bf H}_{ext}\right)^{2} - {\bf E}^{2}\,\right]
        \hspace{0.1in} . \label{hbcs}
\end{eqnarray}
Here $\psi_{\sigma}(x)$ is the field operator for a Landau quasiparticle
with spin $\sigma$; ${\bf H}=\nabla\times {\bf A}$ is the microscopic
magnetic field and its associated vector potential; ${\bf H}_{ext}$ is an
applied magnetic field; ${\bf E}= - (\nabla A_{0} + \partial {\bf A}/
\partial t)$ is the electric field produced by the vortex motion; $g>0$ is the
(attractive) coupling constant; and
repeated indices are summed over. The gap function $\Delta$ is most
conveniently introduced via a Hubbard-Stratonovitch transformation \cite{hs}
which allows eqn.~(\ref{hbcs}) to be written as a path integral over
$(\,\Delta ,\; \Delta^{\ast}\,)$ in which the four Fermion interaction is
replaced by the bilinear Fermion pairing interaction
\begin{equation}
W=\int\, {\cal D}[\Delta ]{\cal D}[\Delta^{\ast}]\,\langle\,vac;
    \Delta (t=T)|U_{\Delta}(T,0)|vac;\Delta(0)\,\rangle \hspace{0.1in} .
\end{equation}
Here $U_{\Delta}(T,0)= {\cal T}(\,\exp [-i\int_{0}^{T}\, d\tau H_{eff}]\,)$,
where $H_{eff}= H_{f} + L_{c} + L_{em}$; and
\begin{eqnarray}
H_{f} & = & \int\, d^{3}x\,\psi^{\dagger}_{\sigma}\left[\, -\frac{1}
               {2}\left(\nabla - ie{\bf A}\right)^{2} -E_{f} +e A_{0}\,\right]
                \psi_{\sigma} \nonumber \\
   & & \mbox{} \hspace{0.25in} +\int\, d^{3}x\, \left[\,\Delta^{\ast}
        \psi_{\downarrow}\psi_{\uparrow}+\Delta\psi^{\dagger}_{\uparrow}
         \psi^{\dagger}_{\downarrow}\, \right]
          \hspace{0.1in} ;
\end{eqnarray}
\begin{equation}
L_{c}+L_{em}= \int\,d^{3}x\,\frac{|\Delta |^{2}}{2g} + \int\, d^{3}x\,
               \frac{1}{8\pi}\left[\,\left(\, {\bf H} - {\bf H}_{ext}\,
                \right)^{2} - {\bf E}^{2} \, \right] \hspace{0.1in} .
\end{equation}
Note that the vacuum state and $H_{eff}$ both depend on the particular
``path'' taken by $\Delta$. As discussed earlier, we are limiting ourselves
to ``paths'' whose time dependence corresponds to adiabatic motion of the
line vortex (i.\ e.\ the vortex velocity satisfies
$|\dot{{\bf r}}_{0}|\ll v_{f}$;
see also Bardeen \cite{bard2}).

The action for the superconducting condensate $S$ is given by
\begin{eqnarray}
e^{-iS} & = & e^{-i\left( S_{0}+S_{hyd} \right) }
     \nonumber \\
  & = & \langle\, vac; \Delta(T)|U_{\Delta}(T,0)|
                            vac;\Delta(0)\, \rangle \nonumber \\
  & = & \langle\, vac; \Delta(T)|U_{\Delta(n)}(t_{n+1},t_{n})\cdots
          U_{\Delta(k)}(t_{k+1},t_{k})\cdots U_{\Delta(0)}(t_{1},t_{0})
           |vac;\Delta(0)\, \rangle \hspace{0.1in} ,
\end{eqnarray}
where $S_{0}$ is the action for the bulk degrees of freedom of the condensate;
$S_{hyd}$ is the action for the hydrodynamic degrees of freedom;
$\Delta(k)\equiv \Delta(t_{k})$; $t_{n+1}=T$; and $t_{0}=0$.
To make use of the Adiabatic Theorem it proves useful to define the
instantaneous eigenstates $|E_{n}(t)\,\rangle$ of the instantaneous
effective Hamiltonian $H_{eff}(t)$. Appropriately inserting complete sets of
instantaneous eigenstates gives
\begin{eqnarray}
e^{-i\left( S_{0}+S_{hyd}\right) }
                & = & \langle\, vac; \Delta(T)|
                                     U_{\Delta(n)}(t_{n+1},t_{n})\times
                                     \nonumber \\
 & & \mbox{} \hspace{0.25in} \left\{ \prod_{k=1}^{n}\sum_{l,m}|E_{l}(t_{k})\,
      \rangle \langle\, E_{l}(t_{k})|U_{\Delta(k)}(t_{k},t_{k-1})
        |E_{m}(t_{k-1})\,\rangle\langle\, E_{m}(t_{k-1})| \right\} \times
          \nonumber \\
 & & \mbox{} \hspace{0.75in} U_{\Delta(0)}(t_{1},t_{0})|vac; \Delta(0)\,
      \rangle \hspace{0.1in} .
\end{eqnarray}
{}From the Adiabatic Theorem, we know that $U_{\Delta(0)}(t_{1},t_{0})$ evolves
$|vac; \Delta(0)\, \rangle$ into an eigenstate of $H_{eff}(t_{1})$
(see Appendix A) which will be denoted $|E_{0}(t_{1});\Delta(t_{1})\,\rangle$.
Similarly, $U_{\Delta(k-1)}(t_{k},t_{k-1})$ evolves $|E_{0}(t_{k-1});
\Delta(t_{k-1})\,\rangle$ into the energy eigenstate of $H_{eff}(t_{k})$
denoted $|E_{0}(t_{k}); \Delta(t_{k})\,\rangle$ for $n\geq k\geq 2$. Thus
\begin{eqnarray}
e^{-i\left( S_{0}+S_{hyd}\right) }
                 & = & \langle\, vac\, ;\Delta(T)|
                                     U_{\Delta(n)}(t_{n+1},t_{n})
                                      |E_{0}(t_{n});\Delta(t_{n})\,\rangle
                                       \times \nonumber \\
 & & \mbox{} \hspace{0.25in} \prod_{k=1}^{n}\langle\, E_{0}(t_{k});
        \Delta(t_{k})|U_{\Delta(k)}(t_{k},t_{k-1})|E_{0}(t_{k-1});
         \Delta(t_{k-1})\, \rangle \times \nonumber \\
 & & \mbox{} \hspace{0.50in} \langle\, E_{0}(t_{1}); \Delta(t_{1})|
      U_{\Delta(0)}(t_{1},t_{0})|vac\, ; \Delta(0)\, \rangle
       \hspace{0.1in} .
    \label{pint}
\end{eqnarray}
{}From the Adiabatic Theorem (see Appendix A), we have that ($\epsilon =
T/(n+1)$)
\begin{eqnarray}
\lefteqn{\mbox{} \hspace{-1in}\langle\, E_{0}(t_{k}); \Delta(k)|
   U_{\Delta(k)}(t_{k},t_{k-1})|
     E_{0}(t_{k-1}); \Delta(k-1)\, \rangle} \nonumber \\
 & = & \langle\, E_{0}(t_{k});\Delta(k)|e^{i\dot{\gamma}_{B}\epsilon}
        e^{-iE_{0}(t_{k})\epsilon}|E_{0}(t_{k}); \Delta(k)\,
         \rangle \nonumber \\
 & = & e^{i\dot{\gamma}_{B}\epsilon}\langle\, E_{0}(t_{k});\Delta(k)|
        e^{-iH_{eff}(t_{k})\epsilon}|E_{0}(t_{k});
         \Delta(k)\,\rangle \hspace{0.1in} . \label{matr}
\end{eqnarray}
Here $\dot{\gamma}_{B}$ is the time derivative of the gauge invariant
generalization of the Berry phase at time
$t_{k}$. The matrix element appearing in the final line of eqn.~(\ref{matr})
can be written as a path integral over the fermionic degrees of freedom
which can be evaluated exactly since $H_{f}$ is quadratic in the fermion
field operators. This calculation is described in Appendix B and the result
is given by $\exp [\,-i\tilde{S}\,]\,$\footnote{The result quoted in
eqn.~(\ref{action}) does not include the contribution $S_{1}$ found in
Appendix B. This is because we will obtain $S_{1}$ in this subsection from
the ground state Berry phase $\Gamma$ (found in Section~\ref{sec2b}).}, where
\begin{eqnarray}
\tilde{S} & = & S_{0} -\int\,d^{3}x\,\frac{|\Delta |^{2}}{2g} +\int\, d^{3}x
                 d\tau\,\left[ \,\frac{m}{2}\rho_{s}
         {\bf v}_{s}^{2}({\bf x},\; t) + N(0)\tilde{A}_{0}^{2}\, \right]
           \hspace{0.1in} .
\label{action}
\end{eqnarray}
Here $S_{0}$ is the action for the bulk degrees of freedom, which is not of
immediate interest to us, and so will not be written out explicitly;
$\rho_{s}$ is the
density of superconducting electrons at $T=0$; ${\bf v}_{s}=-(h/2m)[ \nabla
\phi + (2e{\bf A})/(\hbar c) ]$; $\phi$ is the phase of the gap function;
$N(0)$ is the electron density of states at the Fermi level; $\tilde{A}_{0}
=eA_{0} + (\hbar /2)\partial_{t}\phi$; and $\hbar$, $m$ and $c$ have been
re-instated. Thus from eqns.~(\ref{pint}),
(\ref{matr}) and (\ref{action}), the
hydrodynamic action is
\begin{equation}
S_{hyd}=\int\, d\tau \, \left[\, - \hbar\dot{\gamma}_{B} +
           \int\, d^{3}x\,\left[\,\frac{m\rho_{s}}{2}{\bf v}_{s}^{2} +
            N(0)\tilde{A}_{0}^{2} +
            \frac{1}{8\pi}\left\{ \left({\bf H}-{\bf H}_{ext}\right)^{2}
             -{\bf E}^{2} \right\}
             \right]\:\right]\hspace{0.1in} .
\end{equation}
We see that the Berry phase induced in the superconducting ground state
by the vortex motion has found its way into $S_{hyd}$. Such contributions
to low energy effective actions by Berry phases are well-known and the
induced term is referred to as a Wess-Zumino term \cite{wz}.
Wess-Zumino terms are topological in origin and in our case it is
the vortex topology which is responsible for the occurrence of such a term
in $S_{hyd}$.
We are interested in a superflow which is the sum of an applied steady
superflow ${\bf v}= (\hbar /2m)\nabla\beta$ and one that circulates
with velocity ${\bf v}_{circ}({\bf r};\; {\bf r}_{0}(t))$ about
the adiabatically moving vortex. The gap phase acts as the velocity potential
for the superflow $\phi = \beta -\theta({\bf r};\; {\bf r}_{0}(t))$, and
$\theta$ is the azimuthal angle about the vortex at ${\bf r}_{0}(t)$.

In the absence of pinning centers, the energy of the ground state will not
depend on the location of the vortex so that the initial ground state
$|vac\, ;\Delta(0)\,\rangle$ will evolve adiabatically into the ground state
of $H_{eff}(t)$ for all $t$ of interest. Thus $\gamma_{B}$ is the ground state
Berry phase $\Gamma$ calculated in Section~\ref{sec2b}. Making use of
eqn.~(\ref{bigres}) gives
\begin{equation}
S_{hyd}  =  \int\,d\tau\,d^{3}x\left[\,
               \begin{array}{l}
                 \frac{\rho_{s}\hbar}{2}\left(\,
                  \dot{{\bf r}}_{0}\cdot\nabla_{{\bf r}_{0}}\theta -\frac{2e}
                   {\hbar} A_{0}\,\right) \\
                 {} \hspace{0.25in} + \frac{m\rho_{s}}{2}{\bf v}_{s}^{2} +
                  N(0)\tilde{A}_{0}^{2}  \\
                 {} \hspace{0.5in} +
                  \frac{1}{8\pi}\left\{\, \left({\bf H}-{\bf H}_{ext}
                   \right)^{2}\, -{\bf E}^{2}\, \right\}
                \end{array} \right] \hspace{0.1in} .
              \label{act}
\end{equation}
The terms in $S_{hyd}$ linear in $\nabla_{{\bf r}_{0}}\theta$ describe the
coupling of the vortex to the applied superflow ${\bf v}$; to the
electric and magnetic field via ($A_{0}$, ${\bf A}$); and to the
superconducting
condensate via the Berry phase $\Gamma$. Thus we re-write eqn.~(\ref{act}) as
\begin{eqnarray}
S_{hyd} & = & \int\, d\tau\, d^{3}x\left[ \, {\cal L}_{cond}
                 -\frac{\rho_{s}\hbar}{2}\nabla_{{\bf r}_{0}}\theta\cdot
                   \left( {\bf v}-\frac{e}{mc}{\bf A}+
                    2eA_{0}\frac{N(0)}{\rho_{s}}\dot{{\bf r}}_{0} -
                    \dot{{\bf r}}_{0}
                    \right)\,\right] \nonumber \\
 & = & S_{cond} + S_{int} \hspace{0.1in} . \label{eact}
\end{eqnarray}
The non-dissipative force acting on the vortex is given by the functional
derivative of $S_{int}$ with respect to ${\bf r}_{0}$. Carrying out the
functional derivative gives
\begin{equation}
{\bf F}_{nd}=-\frac{\rho_{s}h}{2}\left( {\bf v}-\dot{{\bf r}_{0}}\right)
               \times\hat{{\bf z}} + {\cal O}\left( \frac{\xi^{2}}{\lambda^{2}}
                \right)
\end{equation}
for a vortex with winding number $\omega = -1$, which is the case we have
been considering. The effect of the scalar and vector potential on the vortex
is seen
to contribute to higher order in $\xi_{0}^{2}/\lambda^{2}$, where $\xi_{0}$ is
the zero temperature coherence length and $\lambda$ is the London penetration
depth. For arbitrary winding number, we have the general result
\begin{equation}
{\bf F}_{nd}=\frac{\rho_{s}h\omega}{2}\left( {\bf v}-\dot{{\bf r}}_{0}\right)
   \times\hat{{\bf z}} \hspace{0.1in} .
   \label{force}
\end{equation}
This is the principal result for Section~\ref{sec2}  and is seen to agree with
the result of Ao and Thouless \cite{ath} which was based on an ansatz for the
many-body ground state. In this section we have worked with the {\em true\/}
superconducting ground state (in the presence of a vortex) and found that
the Berry phase generated in this ground state {\em is\/} responsible for
producing a Magnus (lift) force on the vortex as argued by Ao and Thouless
\cite{ath}, and that ${\bf F}_{nd}$ is
given by the NV-form. In the following section
we will focus on the momentum flows occuring in the superconductor and
from this analysis we will be able to determine the non-dissipative force on
the vortex. The result obtained in Section~\ref{sec3}
will be found to be identical to the result in eqn.~(\ref{force}).

\section{Linear Momentum Flow Analysis and ${\bf F}_{nd}$}
\label{sec3}
In this section we will examine the flow of linear momentum in a superconductor
when a moving line vortex is present. Our microscopic analysis will be
based on the Bogoliubov equation. In the first subsection we will re-write
the Bogoliubov dynamics in a pseudo-relativistic notation which proves
convenient for calculational purposes. In Section~\ref{sec3b} we work out
the continuity equation for the condensate linear momentum at $T=0$. This
equation is seen to contain a source term which is shown in Section~\ref{sec3c}
to be non-vanishing and follows from the non-invariance of the
measure in a path integral description of the NQP dynamics
under a phase transformation of the NQP field operator.
In Section~\ref{sec3c} we evaluate the
variation of the measure to obtain an explicit expression for the linear
momentum source term. In Section~\ref{sec3d} we show that the continuity
equation is simply the acceleration equation for the superflow. The
topological character of the vortex is seen to lead to a transfer of
linear momentum from the condensate to the vortex which corresponds to the
non-dissipative force ${\bf F}_{nd}$. Our analysis shows that the result
found for ${\bf F}_{nd}$ from tracking the momentum flows agrees exactly with
the result of the Berry phase
calculation. The contribution to ${\bf F}_{nd}$ rooted in the topology of the
vortex and manifesting in the non-invariance of the measure is further seen to
contribute the classical Magnus (lift) force to ${\bf F}_{nd}$.

\subsection{Preliminaries}
\label{sec3a}
Our starting point is the Bogoliubov equation given in eqn.~(\ref{bog}) for
the case where a vortex is present with winding number $\omega = -1$
(viz.\ $\Delta = \Delta_{0}(r)\exp [-i\theta ]$). Substituting
\begin{equation}
\left(\begin{array}{c}
         u^{\prime}_{n} \\
         v^{\prime}_{n}
      \end{array} \right) = e^{\frac{i}{2}\theta\sigma_{3}}
                             \left( \begin{array}{c}
                                       u_{n} \\
                                       v_{n}
                                    \end{array} \right)
\end{equation}
into eqn.~(\ref{bog}) leads to the Bogoliubov Hamiltonian
\begin{eqnarray}
H_{BOG} & = & \left( \begin{array}{cc}
                        \frac{1}{2}\left(i\nabla - {\bf v}_{s}
                         \right)^{2} - E_{f} & \Delta_{0} \\
                        \Delta_{0} & -\frac{1}{2}\left(i\nabla + {\bf v}_{s}
                                       \right)^{2} + E_{f}
                     \end{array} \right) \nonumber \\
 & = & \sigma_{3}\left[ \frac{1}{2}\left(i\nabla - \sigma_{3}{\bf v}_{s}
         \right)^{2} - E_{f}\right] + \Delta_{0}\,\sigma_{1}
          \hspace{0.1in} ,
\end{eqnarray}
where the $\{ \sigma_{i} \} $ are the $2\times 2$ Pauli matrices corresponding
to the two-dimensional Nambu space and have nothing to do with spin; and
${\bf v}_{s}= -(1/2)\nabla\theta - e{\bf A}$ ($\hbar = m = c = 1$).

We now make an eikonal approximation for the eigenstates near the Fermi
surface \cite{eik}
\begin{equation}
\left( \begin{array}{c}
          u_{n}^{\prime} \\
          v_{n}^{\prime}
       \end{array} \right) = e^{i{\bf q}\cdot {\bf r}}
                              \left( \begin{array}{c}
                                        U_{n} \\
                                        V_{n}
                                     \end{array} \right) \hspace{0.1in} ,
\end{equation}
where $|{\bf q}|=k_{f}$ and the ($U_{n}$, $V_{n}$) vary on a
length scale $L\gg k_{f}^{-1}$. To first order in gradients
\begin{equation}
H_{BOG}= \sigma_{3}\left[ -{\bf q}\cdot\left( i\nabla - \sigma_{3}{\bf v}_{s}
          \right)\right] + \Delta_{0}\,\sigma_{1} \hspace{0.1in} .
\end{equation}
This gives rise to the gauge invariant second quantized Lagrangian
\begin{equation}
{\cal L}[\hat{{\bf q}}]= \Psi^{\dagger}\left[ i\partial_{t} +
                          \sigma_{3}\left(\frac{1}{2}\partial_{t}\theta
                           -eA_{0}\right) + \sigma_{3}\, {\bf q}\cdot\left(
                            i\nabla - \sigma_{3}{\bf v}_{s}\right) -
                             \Delta_{0}\,\sigma_{1}\right]\Psi \hspace{0.1in} .
\label{blag}
\end{equation}
We see that the original $3+1$ NQP dynamics has separated into a collection
of independent $1+1$ subsystems labeled by $\hat{{\bf q}}={\bf q}/k_{f}$ which
will be referred to as $\hat{{\bf q}}$-channels. This separation allows us to
reconstruct the $3+1$ dimensional dynamics by adding together the contributions
from the separate $\hat{{\bf q}}$-channels. Because the dynamics of the
$\hat{{\bf q}}$-channels is $1+1$, and the Pauli matrices are the
two-dimensional representation of the Dirac gamma matrices, we can re-write
eqn.~(\ref{blag}) in a pseudo-relativistic notation. With $x^{0}=t$;
$x^{1}= {\bf q}\cdot{\bf x}$; and $\gamma^{0}\equiv \sigma_{1}$,
$\gamma^{1}\equiv -i\sigma_{2}$, $\gamma_{5}\equiv \sigma_{3}$; we can write
\begin{equation}
{\cal L}[\hat{{\bf q}}]= \bar{\Psi}\left[ i\gamma^{\mu}\partial_{\mu} -
       \gamma^{\mu}\gamma_{5}\tilde{A}_{\mu} - \Delta_{0}\right]
        \Psi \hspace{0.1in} .
\end{equation}
Here $\mu = 0,1$; $\tilde{A}_{0}=eA_{0} -(1/2)\partial_{t}\theta$;
$\tilde{A}_{1}= {\bf q}\cdot {\bf v}_{s}$; and $\bar{\Psi}=\Psi^{\dagger}
\gamma^{0}$. Our final manipulation involves analytically continuing to
imaginary time, $x^{0}=-i\bar{x}^{2}$, to obtain the Euclidean Lagrangian
\begin{equation}
{\cal L}_{E}[\hat{{\bf q}}]=\bar{\Psi}\left[\bar{\gamma}^{a}\bar{\partial}_{a}
                             +\bar{\gamma}^{a}\bar{\gamma}_{5}\bar{A}_{a}
                              +\Delta_{0} \right] \Psi \hspace{0.1in} .
\end{equation}
Here $a=1,2$; $d^{2}x = -id^{2}\bar{x}=-id\bar{x}^{2}d\bar{x}^{1}$;
$\tilde{A}_{0}=\bar{A}_{2}$, $\tilde{A}_{1}=-i\bar{A}_{1}$; and $\gamma^{0}
=\bar{\gamma}^{2}$, $\gamma^{1}=i\bar{\gamma}^{1}$, and $\gamma_{5}=
\bar{\gamma}_{5}=-i\bar{\gamma}^{1}\bar{\gamma}^{2}$. These definitions
insure that the operator $i\bar{\gamma}^{a}D_{a}=-i\bar{\gamma}^{a}\left(
\bar{\partial}_{a} + \bar{\gamma}_{5}\bar{A}_{a}\right)$ is Hermitian.

\subsection{The Continuity Equation for the Condensate Linear Momentum}
\label{sec3b}
The result of the last subsection is the separation of the $3+1$ NQP dynamics
into a collection of independent $1+1$ dynamical subsystems to be referred to
as
$\hat{{\bf q}}$-channels. This arises from the eikonal approximation made
for the eigenstates of $H_{BOG}$ near the Fermi surface in terms of wavepackets
with mean momentum $p_{f}\hat{{\bf q}}$. By construction, both positive and
negative energy eigenstates (viz.\ above and below the Fermi surface) carry
mean momentum $p_{f}\hat{{\bf q}}$.

Focusing on the $\hat{{\bf q}}$-channel, its single particle excitations
will be quasiparticles with (mean) momentum $p_{f}\hat{{\bf q}}$ (right go-ers)
and quasi-holes with (mean) momentum $-p_{f}\hat{{\bf q}}$ (left go-ers).
In the superconducting phase, a {\em Landau} quasiparticle with (mean)
momentum $p_{f}\hat{{\bf q}}$ will be a superposition of right go-ers and left
go-ers
\begin{equation}
\psi_{\hat{{\bf q}}}(x)=e^{ip_{f}\hat{{\bf q}}\cdot{\bf x}}\psi_{R}({\bf x};
                         \hat{{\bf q}}) + e^{-ip_{f}\hat{{\bf q}}\cdot
                          {\bf x}}\psi_{L}^{\dagger}({\bf x};\hat{{\bf q}})
      \hspace{0.1in} ,
\end{equation}
where
\begin{eqnarray}
\psi_{R}({\bf x};\hat{{\bf q}}) & = & \sum_{k= -\Lambda}^{\Lambda}
                                       \, e^{ik\hat{{\bf q}}\cdot {\bf x}}
                                        a_{k+k_{f}} \nonumber \\
\psi^{\dagger}_{L}({\bf x};\hat{{\bf q}}) & = & \sum_{k=-\Lambda}^{\Lambda}\,
                                   e^{-ik\hat{{\bf q}}\cdot {\bf x}}
                                    a^{\dagger}_{-k-k_{f}} \hspace{0.1in} ;
\end{eqnarray}
$a_{p}$ ($a^{\dagger}_{p}$) annihalates (creates) a LQP with momentum $p$;
$\Lambda$ is a cutoff that determines the spread of the wavepacket in momentum
space; and spin indices have been suppressed. From this, the {\em Nambu}
quasiparticle field operator is
\begin{equation}
\Psi_{\hat{{\bf q}}}({\bf x}) = \left( \begin{array}{c}
                                          \psi_{R}({\bf x};\hat{{\bf q}}) \\
                                          \psi^{\dagger}_{L}({\bf x};
                                           \hat{{\bf q}})
                                       \end{array} \right)
\hspace{0.1in} .
\label{fop}
\end{equation}
The Noether current associated with the phase transformation
$\Psi_{\hat{{\bf q}}}\rightarrow
\exp [-i\chi ] \Psi_{\hat{{\bf q}}}$ is
$j^{\mu}(x)=\bar{\Psi}_{\hat{{\bf q}}}\gamma^{\mu}\Psi_{\hat{{\bf q}}}$. Using
eqn.~(\ref{fop}) we see that (suppressing the $\hat{{\bf q}}$-dependence for
the time being)
\begin{eqnarray}
j_{0}(x) & \equiv & \lim_{\epsilon\rightarrow 0^{+}}\Psi^{\dagger}(x +
                     \epsilon )\Psi (x) \nonumber \\
 & = & \lim_{\epsilon\rightarrow 0^{+}}\left[\,\psi^{\dagger}_{R}(x+\epsilon )
        \psi_{R}(x)+\psi_{L}(x+\epsilon )\psi^{\dagger}_{L}(x) \,\right]
         \nonumber \\
 & = & \psi^{\dagger}_{R}(x)\psi_{R}(x) - \psi^{\dagger}_{L}(x)\psi_{L}(x)
        \hspace{0.1in} .
\end{eqnarray}
Clearly, $j_{0}(x)$ is the net Fermion number density (particle minus hole
density) and $p_{f}\hat{{\bf q}}\, j_{0}(x)$ is the operator representation of
the density of linear momentum along $\hat{{\bf q}}$
\begin{equation}
{\bf g}(x;\hat{{\bf q}})= p_{f}\hat{{\bf q}}\, j_{0}(x) \hspace{0.1in} .
\label{mom}
\end{equation}
Taking the vacuum expectation value of eqn.~(\ref{mom}) and summing over all
$\hat{{\bf q}}$-channels gives the density of linear momentum in the
condensate
\begin{equation}
g_{i}({\bf x})= k_{f}^{3}\sum_{\alpha}\,\int\,\frac{d^{2}\hat{{\bf q}}}
  {4\pi^{2}}\,\hat{{\bf q}}_{i}\langle\, vac|j^{0}(x;\hat{{\bf q}})|vac\,
   \rangle_{\hat{{\bf q}}} \hspace{0.1in} .
\label{momden}
\end{equation}
Here $\alpha$ is the spin index $(\pm)$ and $|vac\,\rangle_{\hat{{\bf q}}}$
is the vacuum (ground) state for the $\hat{{\bf q}}$-channel. Similarly, the
condensate stress tensor is
\begin{equation}
T_{ij}=k_{f}^{3}\sum_{\alpha}\,\int\,\frac{d^{2}\hat{{\bf q}}}{4\pi^{2}}\,
   \hat{{\bf q}}_{i}\hat{{\bf q}}_{j}\,\langle\, vac |j^{1}(x;\hat{{\bf q}})|
    vac\, \rangle_{\hat{{\bf q}}} \hspace{0.1in} .
\label{stress}
\end{equation}
The continuity equation for the condensate linear momentum is then
\begin{eqnarray}
\partial_{t}g_{i} + \partial_{j}T_{ij} & = & k_{f}^{3}\sum_{\alpha}\,\int\,
                                          \frac{d^{2}\hat{{\bf q}}}{4\pi^{2}}
                                           \,\hat{{\bf q}}_{i}\langle\, vac|
                                            \partial_{t}j^{0}+\hat{{\bf q}}_{j}
                                             \partial_{j}j^{1}|vac \,
                                              \rangle_{\hat{{\bf q}}}
                                               \nonumber \\
 & = & k^{3}_{f}\sum_{\alpha}\,\int\,\frac{d^{2}\hat{{\bf q}}}{4\pi^{2}}\,
        \hat{{\bf q}}_{i}\,\langle\, vac|\partial_{\mu}j^{\mu}|vac \,
         \rangle_{\hat{{\bf q}}} \hspace{0.1in} .
\label{con2}
\end{eqnarray}
We go on to evaluate the matrix element appearing in eqn.~(\ref{con2}) and will
see that it is non-vanishing due to the non-invariance of the measure in the
path integral used to formulate the NQP dynamics.

\subsection{Non-invariance of the Measure}
\label{sec3c}
The generating functional for the NQP Green's functions in the
$\hat{{\bf q}}$-channel is
\begin{equation}
W_{NQP}\left[\,\eta,\bar{\eta};\hat{{\bf q}}\right] =
  {\cal N}\int\, {\cal D}\left[ \Psi \right]\,\exp \left[\,i\int\, d^{4}x
   \left\{\,\bar{\Psi}\left(\, i\gamma^{\mu}\partial_{\mu} -\gamma^{\mu}
    \gamma_{5}A_{\mu} - \Delta_{0}\,\right)\Psi
     +\bar{\eta}\Psi + \bar{\Psi}\eta\,\right\}\,\right]
     \hspace{0.1in} .
\end{equation}
Here $\eta$ ($\bar{\eta}$) are external sources for $\bar{\Psi}$ ($\Psi$);
and ${\cal N}$ is a normalization factor.
Invariance of $W_{NQP}$ under the change of variable $\Psi (x)\rightarrow
\exp [-i\theta (x)]\Psi (x)$ gives
\begin{eqnarray}
0 & = & \int\, {\cal D}\,\left[\Psi\right]\,\int\, d^{4}y\, \exp\left[\,
         i\int\, d^{4}x\,\left\{\,\bar{\Psi}\left(i\gamma^{\mu}\partial_{\mu}
          - \gamma^{\mu}\gamma_{5}A_{\mu} - \Delta_{0}\right)\Psi +
           \bar{\eta}\Psi + \bar{\Psi}\eta\,\right\}\,\right]
            \nonumber \\
 & & \mbox{} \hspace{0.5in} \times\theta (y)\left[\, -i\partial_{\mu}j^{\mu}
       +iF(y) + \bar{\eta}\Psi + \bar{\Psi}\eta\,\right] \hspace{0.1in} .
\label{change}
\end{eqnarray}
$F(y)$ enters eqn.~(\ref{change}) through the Jacobian of the transformation
$d\left[\Psi^{\prime}\right]=J\, d\left[\Psi\right]$ and (as we shall see
shortly), $J$ has the form: $J=\exp \left[\, i\int\, d^{4}x\,\theta(x)F(x)
\,\right]$. Setting $\eta =\bar{\eta} = 0$ in eqn.~(\ref{change}) gives
\begin{equation}
\partial_{\mu}j^{\mu} = F(x) \hspace{0.1in} .
\end{equation}

To calculate the Jacobian we follow Fujikawa \cite{fuj} and go over to
Euclidean space $x^{0}\rightarrow -i\bar{x}^{2}$. From Section~\ref{sec3a},
this transformation leads to the Euclidean Lagrangian
\begin{eqnarray}
{\cal L}_{E}\left(\hat{{\bf q}}\right) & = & -\bar{\Psi}\left(\,
                                              \bar{\gamma}^{a}
                                               \bar{\partial}_{a} +
                                                \bar{\gamma}^{a}
                                                 \bar{\gamma}_{5}\bar{A}_{a} +
                                                  \Delta_{0}\,\right)\Psi
                                                   \nonumber \\
 & = & \bar{\Psi}\left(\,\bar{\gamma}^{a}D_{a} - \Delta_{0}\,\right)\Psi
         \hspace{0.1in} ,
\end{eqnarray}
and
\begin{equation}
iS\rightarrow -S_{E} = - \int\, d^{4}x_{E}\,\bar{\Psi}\left(\,
  \bar{\gamma}^{a}\bar{\partial}_{a} + \bar{\gamma}^{a}\bar{\gamma}_{5}
   \bar{A}_{a} + \Delta_{0}\,\right) \Psi \hspace{0.1in} .
\end{equation}
The continuation to Euclidean space has been done in such a manner as to insure
that $i\bar{\gamma}_{a}D_{a}$ is Hermitian and so has a complete set of
states $i\bar{\gamma}^{a}D_{a}\phi_{n} = \Lambda_{n}\phi_{n}$.

Under the change of variable $\Psi\rightarrow\exp \left[ -i\theta(x)\right]
\Psi$,
\begin{equation}
\Psi\rightarrow\Psi^{\prime} = \sum_{n}\, a_{n}^{\prime}\phi_{n} =
  \sum_{n}\, a_{n}e^{-i\theta}\phi_{n} \hspace{0.1in} ,
\end{equation}
where the field operator has been expanded in the complete set $\{\phi_{n}\}$.
Thus
\begin{equation}
a^{\prime}_{n}=C_{nm}a_{m}=e^{-i\theta }a_{n} \hspace{0.25in}
 \Longrightarrow \hspace{0.25in} C_{nm}=\delta_{nm}e^{-i\theta}
  \hspace{0.1in} .
\end{equation}
As $\Psi$ is a Grassmann variable
\begin{eqnarray}
J = & \left( {\rm det}\, C \right)^{-1} & = \exp\left[\, i\int\, d^{4}x_{E}\,
                                           \sum_{n}\,\phi^{\dagger}_{n}(x)
                                            \theta (x) \phi_{n}(x)\,\right]
                                             \nonumber \\
 & & = \exp\left[\, i\int\, d^{4}x_{E}\,\theta (x)F_{E}(x)\,\right]
       \hspace{0.1in}
\end{eqnarray}
as anticipated, and
\begin{equation}
F_{E}(x)= \sum_{n}\,\phi^{\dagger}_{n}(x)\phi_{n}(x)= \sum_{n}\, {\rm tr}
  \phi_{n}(x)\phi^{\dagger}_{n}(x) \hspace{0.1in} .
\end{equation}
Here tr is a sum over Nambu indices only. $F_{E}(x)$ is divergent and its
finite part is isolated by introducing a cutoff $M$ and a subtraction to
remove the divergent part. Thus
\begin{equation}
F_{E}(x) = \lim_{M^{2}\rightarrow \infty}\left[\,\sum_{n}\,\phi^{\dagger}_{n}
 (x)e^{-\frac{\Lambda^{2}}{M^{2}}}\phi_{n}(x) - \sum_{m}\,
  \varphi^{\dagger}_{m}(x)e^{-\frac{\lambda^{2}}{M^{2}}}\varphi_{m}(x)\,
   \right] \hspace{0.1in} .
\end{equation}
The $\{\varphi\}$ are eigenfunctions of $i\bar{\gamma}^{a}D_{a}|_{\bar{A}=0}
= -i\bar{\gamma}^{a}\bar{\partial}_{a}$ and $\{\lambda_{m}\}$ are the
associated eigenvalues. Then,
\begin{eqnarray}
F_{E}(x) & = & \lim_{M^{2}\rightarrow\infty}\left\{\,\sum_{n}\,
                \phi^{\dagger}_{n}(x)e^{-\frac{(i{\not D})^{2}}{M^{2}}}
                 \phi_{n}(x) - \sum_{m}\,\varphi^{\dagger}_{m}(x)
                  e^{-\frac{(i\bar{\not\partial})^{2}}{M^{2}}}\varphi_{m}(x)
                   \,\right\} \nonumber \\
 & = & \lim_{M^{2}\rightarrow\infty}\lim_{y\rightarrow x}{\rm tr}\left\{
        \, e^{\frac{{\not D}^{2}}{M^{2}}}\sum_{n}\,\phi_{n}(x)
         \phi^{\dagger}_{n}(y) - e^{\frac{\bar{\not\partial}^{2}}{M^{2}}}
          \sum_{m}\,\varphi_{m}(x)\varphi^{\dagger}_{m}(y)\,\right\}
           \nonumber \\
 & = & \lim_{M^{2}\rightarrow\infty}{\rm tr}\int\,\frac{d^{2}k}{4\pi^{2}}\,
        \left\{\, e^{-ik\cdot x}e^{\frac{{\not D}^{2}}{M^{2}}}e^{ik\cdot x}
         -e^{-ik\cdot x}e^{\frac{\bar{\not\partial}^{2}}{M^{2}}}
          e^{ik\cdot x}\,\right\} \nonumber\\
 & = & \lim_{M^{2}\rightarrow\infty}{\rm tr}\int\,\frac{d^{2}k}{4\pi^{2}}\,
        \left\{\,\exp\left[\,\frac{-k_{a}k_{a}+\left(\bar{\partial}_{a}
         \bar{A}_{a}\right)\bar{\gamma}_{5} + \frac{1}{2}\bar{\gamma}_{5}
          \bar{\gamma}^{a}\bar{\gamma}^{b}\bar{F}_{ab}}{M^{2}}\,\right]
           -\exp\left[\,\frac{-k_{a}k_{a}}{M^{2}}\,\right]\,\right\}
            \nonumber\\
 & = & \lim_{M^{2}\rightarrow\infty}\frac{M^{2}}{4\pi}{\rm tr}\left[\,
        \frac{\bar{\gamma}_{5}\left(\bar{\partial}_{a}\bar{A}_{a}\right) +
               \frac{1}{2}\bar{\gamma}_{5}\bar{\gamma}^{a}\bar{\gamma}^{b}
                \bar{F}_{ab}}{M^{2}} + {\cal O}\left(\frac{1}{M^{4}}\right)
                 \,\right] \nonumber\\
 & = & \frac{i}{4\pi}\epsilon^{ab}\bar{F}_{ab} \hspace{0.1in} ,
\end{eqnarray}
where $\epsilon^{01}\equiv 1$. In going from the second to the third line
the completeness of the $\{\phi_{n}\}$ ($\{\varphi_{m}\}$) has been used as
well
as the $1+1$ character of the $\hat{{\bf q}}$-channel dynamics.

Switching back to Minkowski space gives $F(x)= (\epsilon^{\mu\nu}
\tilde{F}_{\mu\nu})/4\pi$ so that
\begin{equation}
\partial_{\mu}j^{\mu} = \frac{\epsilon^{\mu\nu}\tilde{F}_{\mu\nu}}{4\pi}
  \hspace{0.1in} .
\label{anomal}
\end{equation}
In the following subsection we will use this result to obtain the acceleration
equation for the condensate superflow.

\subsection{The Acceleration Equation and ${\bf F}_{nd}$}
\label{sec3d}
Inserting eq.~(\ref{anomal}) into eqn.~(\ref{con2}), the continuity equation
becomes
\begin{eqnarray}
\partial_{t}g_{i} + \partial_{j}T_{ij} & = &
           k^{3}_{f}\,\sum_{\alpha}\,\int\,\frac{d^{2}\hat{{\bf
q}}}{4\pi^{2}}\,
            \hat{{\bf q}}_{i}\frac{1}{2\pi}\tilde{F}_{01} \nonumber\\
 & = & k^{3}_{f}\,\sum_{\alpha}\,\int\,\frac{d^{2}\hat{{\bf q}}}{8\pi^{3}}\,
             \hat{{\bf q}}_{i}\left[\,\partial_{0}\tilde{A}_{1} - \partial_{1}
              \tilde{A}_{0}\,\right] \nonumber\\
 & = & k^{3}_{f}\,\sum_{\alpha}\,\int\,\frac{d^{2}\hat{{\bf q}}}{8\pi^{3}}\,
        \hat{{\bf q}}_{i}\hat{{\bf q}}_{j}\left[\,\partial_{0}
         \left(-\frac{1}{2}\partial_{j}\theta - eA_{j}\right) - \partial_{j}
          \left(eA_{0}-\frac{1}{2}\partial_{t}\theta\right)\,\right]
           \nonumber\\
 & = &  {\rm C}_{0}\left\{\, -\frac{\hbar}{2}\left[\,\partial_{0},\;
         \partial_{j}\,\right]\theta + eE_{j}\,\right\} \hspace{0.1in} ,
\label{acceqn}
\end{eqnarray}
where ${\rm C}_{0}= k^{3}_{f}/3\pi^{2}$ is the particle density in the
normal phase when the chemical potential equals the Fermi energy; and $\hbar$
has been restored. The source term is clearly non-vanishing and contains the
effects of the electric field whose origin is the vortex motion as well as a
term whose origin is the topology of the vortex. The essential topological
property of the
vortex is that the phase of the gap function changes by $2\pi\omega$ as we
wind once around the vortex (recall $\omega$ is the vortex winding number)
so that
\begin{equation}
\left[\,\partial_{x},\;\partial_{y}\,\right]\phi = 2\pi\omega\,\delta
   (x-x_{0})\delta(y-y_{0}) \hspace{0.1in} ,
\label{top1}
\end{equation}
where $\phi$ is the gap phase ($\phi = -\theta$ for our calculation). Thus
\begin{equation}
-\frac{\hbar}{2}\left[\,\partial_{0},\;\partial_{j}\,\right]\theta =
 \frac{h\omega}{2}\left( \dot{{\bf r}}_{0}\times\hat{{\bf z}}\right)
  \delta^{2}({\bf r}-{\bf r}_{0}) \hspace{0.1in} .
\label{top2}
\end{equation}
We also note that since ${\bf v}_{s}= -(\hbar\nabla\theta)/2 -e{\bf A}$,
\begin{equation}
{\bf v}_{s}\times (\nabla\times {\bf v}_{s}) + \frac{\hbar}{2}{\bf v}_{s}
 \times(\nabla\times\nabla\theta ) + e{\bf v}_{s}\times{\bf B} = 0
  \hspace{0.1in} .
\end{equation}
It is straightforward to show that (recall $\phi = -\theta$)
\begin{equation}
{\bf v}_{s}\times (\nabla\times\nabla\theta ) = -\frac{h\omega}{2}
  \left( {\bf v}_{s}\times\hat{{\bf z}}\right)\delta^{2}({\bf r}-{\bf r}_{0})
 \hspace{0.1in} ,
\label{top3}
\end{equation}
so that
\begin{equation}
{\bf v}_{s}\times(\nabla\times{\bf v}_{s}) - \frac{h\omega}{2}
  \left({\bf v}_{s}
  \times\hat{{\bf z}}\right)\delta^{2}({\bf r}-{\bf r}_{0}) + e{\bf
v}_{s}\times
   {\bf B} = 0 \hspace{0.1in} .
\label{top44}
\end{equation}
Eqn.~(\ref{top44}) is recognized as the (cross product of ${\bf v}_{s}$ with
the) London equation in the presence of a
vortex \cite{f+h}, and follows automatically from our definition of
${\bf v}_{s}$.
Thus eqn.~(\ref{acceqn}) can be re-written as
\begin{equation}
\partial_{t}g_{i} + \partial_{j}T_{ij} = {\rm C}_{0}\left[\, {\bf v}_{s}\times
  (\nabla\times{\bf v}_{s}) + e{\bf E} + e{\bf v}_{s}\times {\bf B}-
   \frac{h\omega}{2}({\bf v}_{s}-\dot{{\bf r}}_{0})\times\hat{{\bf z}}
    \,\delta^{2}({\bf r}-{\bf r}_{0})\,\right]_{i} \hspace{0.1in} .
\label{acc2}
\end{equation}
It is possible to evaluate the LHS of eqn.~(\ref{acc2}) using
eqns.~(\ref{momden})
and (\ref{stress}) once the matrix element
$\langle\, vac|j^{\mu}|vac \,\rangle_{\hat{{\bf q}}}$ is known. A simple
diagrammatic calculation along the lines of Goldstone and Wilczek \cite{gw}
gives
\begin{equation}
\langle\, vac|j^{\mu}|vac \,\rangle_{\hat{{\bf q}}}= \frac{\epsilon^{\mu\nu}}
 {2\pi}\tilde{A}_{\nu} \hspace{0.1in} .
\label{element}
\end{equation}
We will not reproduce that calculation here as eqn.~(\ref{element}) is clearly
consistent with eqn.~(\ref{anomal}) (though see ref.~\onlinecite{fg2} if
further
details are desired). Thus
\begin{eqnarray}
g_{i} & = & k^{3}_{f}\sum_{\alpha}\int\,\frac{d^{2}\hat{{\bf q}}}{4\pi^{2}}\,
             \hat{{\bf q}}_{i}\frac{1}{2\pi}\tilde{A}_{1} \nonumber\\
  & = & k^{3}_{f}\sum_{\alpha}\int\,\frac{d^{2}\hat{{\bf q}}}{8\pi^{3}}
         \hat{{\bf q}}_{i}\hat{{\bf q}}_{j}({\bf v}_{s})_{j} \nonumber\\
  & = & \frac{k^{3}_{f}}{3\pi^{2}}({\bf v}_{s})_{i} \nonumber\\
  & = & {\rm C}_{0}({\bf v}_{s})_{i} \hspace{0.1in} ,
\label{gi}
\end{eqnarray}
and
\begin{eqnarray}
T_{ij} & = & k^{3}_{f}\sum_{\alpha}\int\,\frac{d^{2}\hat{{\bf q}}}{4\pi^{2}}\,
              \hat{{\bf q}}_{i}\hat{{\bf q}}_{j}\,\left( -\frac{1}{2\pi}\right)
               \tilde{A}_{0} \nonumber\\
       & = & {\rm C}_{0}\left(\frac{\hbar}{2}\partial_{t}\theta - eA_{0}\right)
              \delta_{ij} \hspace{0.1in} .
\label{tij}
\end{eqnarray}
Combining eqns.~(\ref{acc2}), (\ref{gi}), and (\ref{tij}); together with some
algebra gives
\begin{eqnarray}
\lefteqn{\frac{\partial}{\partial t}{\bf v}_{s} -{\bf v}_{s}\times(\nabla\times
  {\bf v}_{s}) = } \\
 & & \mbox{} \hspace{0.5in}\nabla (eA_{0})-\nabla (\frac{\hbar}{2}
       \partial_{t}\theta )
   + e{\bf E} + e{\bf v}_{s}\times{\bf B} -\frac{h\omega}{2}({\bf v}_{s}
    -\dot{{\bf r}}_{0})\times\hat{{\bf z}}\,\delta^{2}({\bf r}-{\bf r}_{0})
     \hspace{0.1in} .
\end{eqnarray}
{}From the Josephson equation $(\hbar\partial_{t}\phi )= -\mu_{0}$, where
$\mu_{0}$ is the chemical potential in the vortex rest frame and can be
written as $\mu_{0}=\mu + {\bf v}_{s}^{2}/2 + eA_{0}$ ($\mu$ is the
chemical potential in the lattice rest frame and $m=1$). Recalling that
$\phi = -\theta$, together with the preceeding remarks gives
\begin{equation}
\frac{\partial}{\partial t}{\bf v}_{s} + \nabla (\frac{1}{2}{\bf v}_{s}^{2})
  -{\bf v}_{s}\times (\nabla\times{\bf v}_{s}) = -\nabla\mu +e{\bf E} +
   e{\bf v}_{s}\times{\bf B} -\frac{h\omega}{2}\left( {\bf v}_{s} -
    \dot{{\bf r}}_{0}\right)\times\hat{{\bf z}}\,\delta^{2}({\bf r}-{\bf
r}_{0})
     \hspace{0.1in} ,
\end{equation}
or finally,
\begin{equation}
\frac{d{\bf v}_{s}}{dt}=-\nabla\mu + e{\bf E} +e{\bf v}_{s}\times {\bf B} -
 \frac{h\omega}{2}\left({\bf v}_{s} - \dot{{\bf r}}_{0}\right)\times
  \hat{{\bf z}}\,\delta^{2}({\bf r}-{\bf r}_{0}) \hspace{0.1in} ,
\label{nvres}
\end{equation}
which is the acceleration equation for the condensate superflow. Thus we find
the expected forces related to the hydrodynamic pressure ($\nabla P =
\rho_{s} \nabla
\mu$), and the electric and magnetic fields. We also see that linear
momentum (recall $m=1$) is disappearing from the condensate into the vortex
at ${\bf r}_{0}(t)$ at the rate $(\rho_{s} h\omega /2)({\bf v}_{s}-
\dot{{\bf r}}_{0})\times\hat{{\bf z}}$ per unit length so that
\begin{equation}
{\bf F}_{nd} = \frac{\rho_{s} h\omega}{2}\left( {\bf v}_{s}-\dot{{\bf r}}_{0}
  \right)\times\hat{{\bf z}} \hspace{0.1in} ,
\end{equation}
in agreement with our Berry phase calculation. It is clear from
eqns.~(\ref{top1}), (\ref{top2}) and (\ref{top3}) that the origin of this
force is
the vortex topology and that the Magnus force contribution to ${\bf F}_{nd}$
arises from the non-invariance of the path integral measure. Our result is
also consistent with Ref.~\onlinecite{nv} which showed that the first 3 terms
on the RHS of
eqn.~(\ref{nvres}) lead to a flux of linear momentum in towards the vortex at
the rate $(\rho_{s}h\omega /2)({\bf v}_{s}-\dot{{\bf r}}_{0})\times
\hat{{\bf z}}$
which is exactly equal to the rate at which we find it appearing on the vortex.
This topology driven flow of linear momentum from the $3+1$ dimensional
condensate onto the $1+1$ dimensional vortex is a particular example of the
Callen-Harvey mechanism of anomaly cancellation \cite{chrv} (see below).

\section{Discussion}
\label{sec4}
In this paper we have examined the non-dissipative force ${\bf F}_{nd}$
acting on a line vortex in a type-II superconductor at $T=0$. Our analysis
throughout has been {\em microscopic} and is based on the dynamics inherent
in the Bogoliubov equation. In the first half of this paper we have
verified and strengthened an earlier result of Ao and Thouless \cite{ath}
which argued that ${\bf F}_{nd}$ contains a contribution from the Magnus
(lift) force familiar from classical hydrodynamics and which they argued
was a manifestation of a Berry phase induced in the many-body ground state
by the vortex motion. We have been able to determine the Berry phase induced
in the {\em true\/} superconducting ground state and find that the result
of Ref.~\onlinecite{ath}, which is based on an ansatz for the many-body
ground state wavefunction, agrees with the exact result in the case of a
neutral superfluid. For a charged superconductor, gauge invariance requires
an ``$eA_{0}$''-term to appear in the ground state Berry phase. This
correction is seen however to not effect the final result for ${\bf F}_{nd}$.
It is clear from our calculation that, in the case of a neutral superfluid,
the ground state Berry phase is ultimately a consequence of the topology and
axial symmetry of the vortex. This requires the energy eigenfunctions to be
eigenstates of the z-component of the angular momentum. Examination of the
ansatz of Ref.~\onlinecite{ath} shows that it has captured this essential
property of the true superconducting ground state, and as a consequence, their
calculation produces the exact result in the case of a neutral superfluid.
Although our paper is clearly based
on the BCS theory of superconductivity, the fact that the Berry phase is
ultimately a consequence of the topology and axial symmetry of the vortex
leads us
to suspect that vortices in non-BCS type-II superconductors will likely induce
Berry phases in their corresponding ground state. One would thus expect a
similar Magnus force contribution to ${\bf F}_{nd}$ in non-BCS type-II
superconductors as well.
By deriving the action for the hydrodynamic degrees of freedom of the
superconducting condensate from the microscopic Bogoliubov dynamics we are
able to see how the ground state Berry phase
manifests itself as a Wess-Zumino term in this action \cite{hat}.
Variation of the hydrodynamic action with respect to the vortex
trajectory gives the non-dissipative force acting on the vortex. This
variation also shows that the Wess-Zumino term is responsible
for producing the Magnus force contribution to ${\bf F}_{nd}$. This calculation
substantiates the discussion given by Ao and Thouless which again captures the
essential issues involved. In the second part of this paper we have examined
${\bf F}_{nd}$ by examining microscopically the continuity equation for the
condensate linear momentum. We show that this equation leads to the
acceleration equation for the superflow and by being careful to track the
effects of the vortex topology, are able to see that a flux of linear
momentum flows from the $3+1$ dimensional condensate onto the $1+1$
dimensional vortex where it appears at a rate equal to $(\rho_{s} h\omega /2)
({\bf v}_{s}-\dot{{\bf r}}_{0})\times\hat{{\bf z}}$ (per unit length). The
sink terms in the continuity equation for the condensate linear momentum
are seen to arise from the topology of the vortex. This
topology related transfer of quantum numbers (in our case, linear momentum)
from a higher dimensional system $S_{1}$ in which a lower dimensional system
$S_{2}$ is embedded is an example of the Callen-Harvey mechanism of anomaly
cancellation \cite{chrv}.
Specifically, linear momentum is not conserved in either the condensate or
the vortex, but is conserved in the combined system. In our case, the issue
of anomaly cancellation simply describes the linear momentum transfers
expected between the condensate-vortex subsystems. It should be stressed that
our calculational procedure is able to keep track of the more subtle
momentum transfers that are a consequence of the vortex topology.
Both calculations yield the same result for
the non-dissipative force and verify the presence of the Magnus
force in ${\bf F}_{nd}$ which is seen to be a manifestation of the vortex
topology.

It is straightforward to calculated the Berry phase for the many-body excited
states. Positive energy NQP excitations are created with $\gamma^{\dagger}_{n
\uparrow}$ and positive energy NQH excitations with
$\gamma^{\dagger}_{n\downarrow}$. For a many-body excited state with NQP's in
the single particle state $(P_{1},\cdots ,P_{N})$ and NQH's in the states
$(H_{1},\cdots ,H_{M})$, the Berry phase is
\begin{eqnarray}
\Gamma_{MB}(P_{1},\cdots ,P_{N};\; H_{1},\cdots ,H_{M}) & = &
           -\sum_{n}\,\phi_{n} + \sum_{\{ P_{i}\} }\,\phi_{P_{i}}
             +\sum_{\{ H_{i} \} }\,\phi_{H_{i}} \nonumber\\
 & = & \Gamma + \Gamma_{exc} \hspace{0.1in} .
\label{mbb}
\end{eqnarray}
We see that a 2-fluid picture naturally develops in the Berry phase in which
the contribution from the superconducting condensate is given by the ground
state Berry phase $\Gamma$ obtained in Section~\ref{sec2} together with a
contribution
$\Gamma_{exc}$ from the elementary excitations present. Because the Berry phase
gives the orbital angular momentum along the vortex axis present in the
energy eigenstates,
we see that the total orbital angular momentum (along the vortex axis)
in an excited state
receives contributions from the condensate and the elementary excitations.
{}From eqn.~(\ref{orth}) we see that the orbital angular momentum associated
with
the elementary excitations is opposite to that of the condensate so that the
orbital angular momentum of the excited states is less than the
orbital angular momentum in the ground state (viz.\ $\Gamma_{MB} < \Gamma$).
Because experiments are done at finite temperature, it is clearly of
interest to generalize our $T=0$ analysis of ${\bf F}_{nd}$ to finite
temperature. An important question is the effect of temperature on the Berry
phase contribution to the hydrodynamic free energy, and, consequently, of the
effect of temperature on the Magnus force contribution to ${\bf F}_{nd}$.
Naively, one would expect this would lead to a thermal average of the Berry
phase given in eqn.~(\ref{mbb}). As we have just seen, $\Gamma_{MB}< \Gamma$ so
that a thermal average of $\Gamma_{MB}$ is expected to decrease with
increasing temperature. Thus one would naively expect the Magnus force
contribution to ${\bf F}_{nd}$ to correspondingly decrease with temperature.
We should also mention that since $\Gamma$ depends on the superconducting
electron density which decreases and eventually goes to zero with increasing
temperature, whereas
$\Gamma_{exc}$ {\em increases} with increasing temperature,
it becomes clear that the thermal average of
$\Gamma_{MB}$ should change sign as a function of temperature. Because of the
anticipated connection between the thermal average of $\Gamma_{MB}$ and the
Magnus force contribution to ${\bf F}_{nd}$, such a sign change will lead to
a sign change in the Magnus force contribution to ${\bf F}_{nd}$.
Thus ${\bf F}_{nd} = (\rho_{s}h\omega /2)({\bf v}_{s}- \alpha (T)\,
\dot{{\bf r}}_{0})\times\hat{{\bf z}}$, and $\alpha (T)$ changes sign with
temperature. Such a temperature dependence for $\alpha (T)$ will act, in
general, to change the sign of the Hall angle in a Hall effect experiment on a
type-II superconductor in the flux flow regime \cite{lobb}. Such a sign
change in the Hall angle as a function of temperature has been observed
\cite{exper} and the fact that neither the BS or NV models can explain this
sign change is the reason for this effect being known as
the ``sign anomaly in the Hall effect'' in superconductors. Further work on
the temperature dependence of the Berry phase contribution to ${\bf F}_{nd}$
is in progress and is clearly of interest to efforts aimed at understanding
the sign anomaly in the Hall effect. We will report on this work elsewhere.
Volovik \cite{vol} has argued
that, for temperatures satisfying $\Delta\gg k_{B}T\gg (\Delta^{2}/E_{f})$,
thermal broadening of single particle states bound to the vortex
will lead to creation of excitations trapped on the vortex as a consequence of
spectral flow and that this mechanism effectively wipes out the Magnus force
contribution to ${\bf F}_{nd}$ which has been found to be present at $T=0$.
Although Volovik's analysis
is different from ours, it is clear that at these temperatures, the important
excitations for a thermal average of $\Gamma_{MB}$ will come from excitations
bound to the vortex so that the remarks above regarding a weakening of the
Magnus force contribution to ${\bf F}_{nd}$ are qualitatively in agreement
with Volovik's result. It should be mentioned that Volovik does not
discuss the possibility of the Magnus force contribution to ${\bf F}_{nd}$
changing sign with temperature, nor of its possible consequences for
Hall effect experiments on type-II superconductors in the flux flow regime.
Ao et.\ al.\ \cite{ao2} have described a finite temperature calculation of the
non-dissipative force and argue that the $T=0$ form of this force is
preserved for $T\neq 0$ and that the only effect of temperature is to replace
the zero temperature superconducting electron density with its finite
temperature form.

I would like to thank: Dr.\ Ping Ao for interesting me in this problem and for
a number of helpful discussions;
Dr.\ Michael Stone for interesting comments and discussions, particularly
with regard to the issue of gauge invariance;
T. Howell III for constant support; and NSERC of Canada for financial support.

\appendix
\section{The Adiabatic Theorem and Berry's Phase}
\label{appendixA}
In this appendix we provide a derivation of the Adiabatic Theorem being
careful not to overlook the appearance of the geometric/topological phase
discovered by Berry \cite{ber}. Our analysis focuses on the appearance of the
Berry phase in the propogator (viz.\ time development operator) for the
wavefunctions as this is the relevant object for our analysis in
Section~\ref{sec2}.
For a standard treatment of the Adiabatic Theorem along these lines
(minus the Berry phase), see Messiah \cite{mes}.

As in our discussion of Section~\ref{sec2}, we consider a quantum system
coupled to an
environment which is evolving adiabatically. This coupling is described by
the appearance of a set of parameters ${\bf R}(t)=(R_{1}(t),\cdots ,R_{N}(t))$
in the system Hamiltonian $H[\, {\bf R}(t)]$. It proves convenient to
introduce the instantaneous eigenstates of $H[\, {\bf R}(t)]$,
\begin{equation}
H[\, {\bf R}(t)]|E_{j}(t)\,\rangle = E_{j}(t)|E_{j}(t)\,\rangle \hspace{0.1in}
{}.
\end{equation}
Clearly $E_{j}(t)$ and $|E_{j}(t)\,\rangle$ are continuous functions of
${\bf R}(t)$ so that one can introduce an operator $A(t)$ such that
\begin{equation}
|E_{j}(t)\,\rangle = A(t)|E_{j}(0)\,\rangle \hspace{0.1in} ;
  \hspace{0.5in} A(0)\equiv 1   \hspace{0.1in} . \label{cont}
\end{equation}
For simplicity, assume the instantaneous energy eigenvalues are discrete
and non-degenerate throughout $0 \leq t \leq T$. From eqn.~(\ref{cont}),
the projection operator $P_{j}(t)=|E_{j}(t)\,\rangle \langle\, E_{j}(t)|$
obeys
\begin{equation}
P_{j}(t)=A(t)P_{j}(0)A^{\dagger}(t)  \hspace{0.1in} .
\end{equation}
Time evolution is governed by the propogator $U(t,t_{0})$ which satisfies
\begin{equation}
i\hbar\frac{d}{dt}U(t,t_{0}) = H(t)U(t,t_{0}) \hspace{0.1in} .
\end{equation}

We now state the Theorem to be \vspace{0.15in}proved:\\
{\bf Adiabatic Theorem:}\\
If initially the system is in the eigenstate $|E_{j}(0)\,\rangle$
(of $H[\, {\bf R}(0)]$), then the state of the system at time $t$ is given by
the eigenstate of $H[\, {\bf R}(t)]$ that evolves continuously from
$|E_{j}(0)\,\rangle$ (viz.\ $|E_{j}(t)\,\rangle$ to within a phase factor);
or formally,
\begin{equation}
\lim_{t\rightarrow\infty}\,U(t,0)|E_{j}(0)\,\rangle = P_{j}(t)\,
    \lim_{t\rightarrow\infty}\,U(t,0)|E_{j}(0)\,\rangle \hspace{0.1in} .
\end{equation}

\noindent {\bf Proof\vspace{0.15in}:} \\
The operator $A(t)$ is determined by the initial
condition $A(0)=1$ and the differential equation
\begin{equation}
i\hbar\frac{dA}{dt}=K^{\prime}(t)A(t) \hspace{0.1in} ,
\end{equation}
where $K^{\prime}(t)$ must satisfy
\begin{equation}
\left[ \, K^{\prime}(t)\, ,\, P_{j}(t)\,\right] = i\hbar\frac{dP_{j}}{dt}
   \hspace{0.1in} .
\label{diff1}
\end{equation}
Because of the presence of projection operators in eqn.~(\ref{diff1}), this
equation does not determine $K^{\prime}(t)$ uniquely. The most general
solution can be shown to be
\begin{equation}
K^{\prime}(t)=i\hbar\sum_{l}\,\left[\,\frac{dP_{j}}{dt} + P_{l}f_{l}\,
  \right] P_{l} \hspace{0.1in} .
\end{equation}
Here $\{ f_{l}(t)\}$ is an arbitrary collection of operators. Requiring
that the dynamics satisfy the Schrodinger equation will allow us to fix
the $\{ f_{l}(t)\}$.

In the adiabatic limit $U(t,0)\rightarrow A(t)\Phi (t)$ as
$t\rightarrow\infty$,
where
\begin{equation}
\Phi (t) = \sum_{j}\, e^{-i\int_{0}^{t}\,d\tau E_{j}(\tau )}P_{j}(0)
 \hspace{0.1in} .
\end{equation}
In general, $U(t,0)=A(t)\Phi (t)W(t)$, where $W(t)$ is determined by the
condition $W(0)=1$ together with
\begin{equation}
\frac{dW}{dt}=\frac{i}{\hbar}\bar{K}^{\prime}W \hspace{0.1in} ;
\label{diff2}
\end{equation}
and
\begin{equation}
\bar{K}^{\prime}= \Phi^{\dagger}A^{\dagger}K^{\prime}A\Phi \hspace{0.1in} .
\end{equation}
By sandwiching eqn.~(\ref{diff2}) between the states $\langle\,E_{j}(0)|$ and
$|E_{i}(0)\,\rangle$, it can be shown that $W_{ij}\equiv \langle\, E_{j}(0)|
W|E_{i}(0)\,\rangle\sim {\cal O}(1/T)\;\; (i\neq j)$ and so vanishes in the
adiabatic limit. The diagonal elements are easily shown to be given by
\begin{equation}
W_{ii}=\langle \, E_{i}(0)|W|E_{i}(0)\,\rangle = e^{\frac{i}{\hbar}
  \int_{0}^{t}\,d\tau \bar{K}^{\prime}_{ii}} \hspace{0.1in} .
\end{equation}
Thus
\begin{equation}
W=\sum_{i}\, W_{ii}P_{i}(0) \hspace{0.1in} .
\end{equation}
Thus for the situation envisioned in the premise of the Adiabatic Theorem,
\begin{eqnarray}
|\psi(t)\,\rangle & = & U(t,0)|E_{i}(0)\,\rangle \nonumber \\
 & = & A(t)\Phi (t)\sum_{j}\,W_{jj}(t)P_{j}(0)|E_{i}(0)\,\rangle \nonumber \\
 & = & e^{\frac{i}{\hbar}\int_{0}^{t}\, d\tau \left[\,\bar{K}^{\prime}_{ii}-
          E_{i}(\tau )\,\right]} A(t)|E_{i}(0)\,\rangle \nonumber \\
 & = & e^{\frac{i}{\hbar}\int_{0}^{t}\, d\tau \left[\,\bar{K}^{\prime}_{ii}
          -E_{i}(\tau )\,\right]}|E_{i}(t)\,\rangle \hspace{0.1in} .
\end{eqnarray}
This proves the Adiabatic Theorem. To fix the phase, we now determine
$\bar{K}^{\prime}_{ii}$.

By definition,
\begin{eqnarray}
\frac{d}{dt}|E_{j}(t)\,\rangle & = & \frac{d}{dt}A(t)|E_{j}(0)\,\rangle
                                      \nonumber \\
 & = & \frac{1}{i\hbar}K^{\prime}|E_{j}(t)\,\rangle \nonumber \\
 & = & \left[\sum_{l}\left(\frac{dP_{l}}{dt} + P_{l}f_{l}\right)P_{l}\right]
          |E_{j}(t)\,\rangle \nonumber \\
 & = & \left( \frac{dP_{j}}{dt}+P_{j}f_{j}\right) |E_{j}(t)\,\rangle
        \hspace{0.1in} .
   \label{diff3}
\end{eqnarray}
Taking the time derivative of the definition $P_{j}(t)|E_{j}(t)\,\rangle
=|E_{j}(t)\,\rangle $ gives
\begin{equation}
\frac{dP_{j}}{dt}|E_{j}(t)\,\rangle + P_{j}(t)\frac{d}{dt}|E_{j}(t)\,\rangle
  = \frac{d}{dt}|E_{j}(t)\,\rangle \hspace{0.1in} .
\label{diff4}
\end{equation}
Equating eqns.~(\ref{diff3}) and (\ref{diff4}) gives
\begin{equation}
P_{j}f_{j}|E_{j}(t)\,\rangle = P_{j}\frac{d}{dt}|E_{j}(t)\,\rangle
     \hspace{0.1in} .
\end{equation}
{}From this we find
\begin{equation}
\langle\, E_{j}(t)|f_{j}|E_{j}(t)\,\rangle = i\langle\, E_{j}(t)|
   \frac{d}{dt}|E_{j}(t)\,\rangle \hspace{0.1in} ,
\end{equation}
and so
\begin{equation}
\bar{K}^{\prime}_{jj}=i\hbar\left[\,\langle\, E_{j}(t)|\frac{d}{dt}
  |E_{j}(t)\,\rangle +i\hbar\langle\, E_{j}(t)|\frac{dP_{j}}{dt}
   |E_{j}(t)\,\rangle\, \right] \hspace{0.1in} .
\label{kdef}
\end{equation}
But the second term on the RHS of eqn.~(\ref{kdef}) vanishes since
$\frac{d}{dt}\left[\,\langle\, E_{j}(t)|E_{j}(t)\,\rangle\,\right]=0$.
Thus $\bar{K}^{\prime}_{jj}=i\hbar\langle\, E_{j}(t)|\frac{d}{dt}|
E_{j}(t)\,\rangle$ and
\begin{eqnarray}
|\psi(t)\,\rangle & = & e^{-\int_{0}^{t}\, d\tau\langle\, E_{j}|\frac{d}{dt}
                         |E_{j}\,\rangle}e^{-\frac{i}{\hbar}\int_{0}^{t}\,
                          d\tau E_{j}(\tau )}|E_{j}(t)\,\rangle
                           \nonumber \\
 & = & e^{i\gamma_{B}(t)}e^{-\frac{i}{\hbar}\int_{0}^{t}\, d\tau E_{j}(\tau )}
         |E_{j}(t)\, \rangle \hspace{0.1in} .
\end{eqnarray}
Thus we recover the Berry phase contribution to the phase factor appearing
in the Adiabatic Theorem.

The derivation of the Berry phase given above assumes an electrically neutral
system. For a system composed of electrically charged particles such as the
electron gas in a superconductor, the Berry phase must be generalized in a
manner consistent with the gauge symmetry underlying electromagnetism. This
is because the hydrodynamic action $S_{hyd}$ must be gauge invariant, and
the Berry phase induced in the superconducting ground state finds its way into
$S_{hyd}$ (as seen in
Section~\ref{sec2c}). The appropriate generalization has been given by
Aharanov and Anadan \cite{ahar}. We will simply quote their result as the
correct form can be guessed straightforwardly from knowledge of how the Berry
phase transforms under a time dependent gauge transformation,
\begin{equation}
\gamma_{B}(t) = \int_{0}^{t}\, d\tau\,\langle\,E(\tau)| i\frac{d}{d\tau}
                 -\frac{e}{\hbar}A_{0}(\tau)|E(\tau)\,\rangle \hspace{0.1in} .
\label{fnl}
\end{equation}
Eqn.~(\ref{fnl}) will be used in Section~\ref{sec2} of this paper.

\section{The Hydrodynamic Action Revisited}
\label{appendixB}
In this Appendix an alternative derivation of the gauge invariant
Wess-Zumino term will be given. The calculation to be described is
based on Eckern et.\ al.\ \cite{ambe}. We will briefly present their approach
and show how the Wess-Zumino term arises very naturally in this approach.
The result found will be seen to agree with the Berry phase calculation
presented in Section~\ref{sec2c}. It is hoped that the {\em independent\/}
calculation
of the Wess-Zumino term given in this Appendix will reassure the reader that
no sleight-of-hand was perpetrated in the normal ordering of the Berry phase in
Section~\ref{sec2b}. A secondary aim of this Appendix is to point out that the
effective action for a single superconductor obtained in Ref.~\onlinecite{ambe}
also contains a gauge invariant Wess-Zumino term (viz.\ a gauge invariant term
first order in time derivatives of the gap phase).

As in Section~\ref{sec2c}, our starting point is the vacuum-to-vacuum
transition amplitude ($\hbar=m=c=1$)
\begin{equation}
W=\langle\,vac;t=T|\:{\cal T}\left(\, \exp \left[\, -i\int_{0}^{T}\, dt\,
  H_{BCS}(t) \,\right]\,\right)|vac; t=0\,\rangle \hspace{0.1in} ;
\end{equation}
and
\begin{eqnarray}
H_{BCS} & = & \int\, d^{3}x \,\psi^{\dagger}_{\sigma}(x)\left[\, -\frac{1}{2}
               \left(\nabla - ie{\bf A}\right)^{2}-E_{f}+eA_{0}\,\right]
                \psi_{\sigma}(x) \nonumber \\
        &   &  {} \hspace{0.25in} -\frac{g}{2}\int\,d^{3}x\,
                   \psi^{\dagger}_{\sigma}(x)\psi^{\dagger}_{-\sigma}(x)
                    \psi_{-\sigma}(x)\psi_{\sigma}(x) \nonumber \\
        &   &  {} \hspace{0.5in} +\int\, d^{3}x\,\frac{1}{8\pi}\left[\,
                   \left({\bf H}-{\bf H}_{ext}\right)^{2} - {\bf E}^{2}
                    \,\right] \hspace{0.1in} .
\end{eqnarray}
(See Section~\ref{sec2c} for a definition of the symbols.) The quartic
interaction term is removed via a Hubbard-Stratonovitch transformation
\cite{hs} so that $W$ is expressed as a path integral over $(\Delta , \;
\Delta^{\ast})$
\begin{equation}
W=\int\, {\cal D}\left[ \Delta\right]\, {\cal D}\left[ \Delta^{\ast}\right]\,
    \langle\, vac; t=T|\: {\cal T}\, \left(\,\exp\left[\, -i\int_{0}^{T}\,
     dt\, H_{eff}(t)\,\right]\,\right)\, |vac;t=0\,\rangle \hspace{0.1in} ,
\label{matem}
\end{equation}
where
\begin{equation}
H_{eff}=H_{f}+L_{c}+L_{em} \hspace{0.1in} ,
\end{equation}
and
\begin{eqnarray}
H_{f} & = & \int\, d^{3}x\, \psi^{\dagger}_{\sigma}(x)\left[\,\frac{1}{2}
             \left(\nabla - ie{\bf A}\right)^{2} - E_{f} + eA_{0}\,\right]
              \psi_{\sigma}(x) \nonumber \\
      &   &  {} \hspace{0.25in} +\int\, d^{3}x\,
\left[\,\Delta(x)\psi^{\dagger}
               _{\uparrow}(x)\psi^{\dagger}_{\downarrow}(x) + \Delta^{\ast}(x)
                \psi_{\downarrow}(x)\psi_{\uparrow}(x)\,\right]
                  \hspace{0.1in} ;
\end{eqnarray}
\begin{equation}
L_{c}+L_{em} = \int\, d^{3}x\, \frac{|\Delta |^{2}}{2g} +
                \int\, d^{3}x\, \frac{1}{8\pi}\left[\, \left( {\bf H} -
                 {\bf H}_{ext}\right)^{2} - {\bf E}^{2} \,\right]
                  \hspace{0.1in} .
\end{equation}
The condensate effective action $S$ is defined through the matrix element
appearing in eqn.~(\ref{matem})
\begin{equation}
e^{-iS}=e^{-i\int_{0}^{T}\,dt\,\left(L_{c}+L_{em}\right)}\,
         \langle\, vac; t=T| \, {\cal T}\left(\,\exp\left[\, -i\int_{0}^{T}\,
          dt\, H_{f}(t)\,\right]\,\right) |vac;t=0 \,\rangle
           \hspace{0.1in} . \label{hydroac}
\end{equation}
The matrix element in eqn.~(\ref{hydroac}) can be calculated easily once
$H_{f}$ is re-written in terms of the NQP field operator $\Psi$ introduced
in Section~\ref{sec2},
\begin{equation}
H_{f}=\int\, d^{3}x\,d^{3}x^{\prime}\, \Psi^{\dagger}(x) H_{BOG}(x;x^{\prime})
       \Psi(x^{\prime}) \hspace{0.1in} ;
\label{hbogham}
\end{equation}
where
\begin{eqnarray}
H_{BOG}(x;x^{\prime}) & = & \left[\,\left\{\, -\frac{1}{2}\left(\nabla -
                             ie\sigma_{3}\/ {\bf A}\right)^{2} - E_{f}
                              +eA_{0}\,
                              \right\}\sigma_{3} +Re(\Delta)\sigma_{1} -
                               Im(\Delta)\sigma_{2}\,\right] \nonumber \\
                      &   & {} \hspace{0.25in}   \times\delta^{3}({\bf x} -
                                {\bf x}^{\prime})\delta(t- t^{\prime})
\hspace{0.1in} .
\end{eqnarray}
With $H_{f}$ given by eqn.~(\ref{hbogham}),
\begin{equation}
\langle\, vac; t=T|\, {\cal T}\,\left(\,\exp\left[\, -i\int_{0}^{T}\, dt\,
  H_{f}(t)\,\right]\,\right)\,|vac;t=0\,\rangle = \exp\left[\,
  {\rm Tr}\:\ln G^{-1}\,
  \right] \hspace{0.1in} ,
\end{equation}
where
\begin{equation}
G^{-1} = i\delta^{3}({\bf x} - {\bf x}^{\prime})\partial_{t}\delta(t-
          t^{\prime}) - H_{BOG} \hspace{0.1in} .
\end{equation}
As in Section~\ref{sec3a}, we make $Im \Delta = 0$ through the unitary
transformation $H_{BOG} \rightarrow \exp[\,i(\phi/2)\sigma_{3}\,]\:H_{BOG}
\:\exp[\,-i(\phi/2)\sigma_{3}\,]$. This gives
\begin{eqnarray}
G^{-1} & \equiv & G_{0}^{-1} + \delta G^{-1} \nonumber \\
       & = &   \left[\, i\frac{\partial}{\partial t} + \left(\, \frac{1}{2}
                \nabla^{2} + E_{f}\right)\sigma_{3} - \Delta_{0}\sigma_{1}\,
                 \right]\delta^{3}({\bf x} - {\bf x}^{\prime})\delta (t-
                  t^{\prime}) \nonumber \\
       &   & {} \hspace{0.1in} + \left[\, \tilde{A}_{0}\sigma_{3} + i{\bf v}
                _{s}\cdot\nabla - \frac{1}{2}{\bf v}_{s}^{2}\sigma_{3}\,
                 \right] \delta^{3}({\bf x} -{\bf x}^{\prime})\delta(t-
                  t^{\prime}) \hspace{0.1in} .
\end{eqnarray}
Here $\tilde{A}_{0}$ and ${\bf v}_{s}$ are defined as in Section~\ref{sec3a}
$(\phi = - \theta)$; and $\delta G^{-1} \equiv \delta G^{-1}_{t1} +
\delta G^{-1}_{s1} + \delta G^{-1}_{s2}$ contains 3 contributions which are
gauge invariant and are (respectively) first order in time derivatives of the
gap phase, first order in space derivatives (of the gap phase), and second
order in space derivatives. Putting these results together gives
\begin{equation}
S= i\, {\rm Tr}\,\ln G^{-1} + S_{c} + S_{em} \hspace{0.1in} ,
\label{treqn}
\end{equation}
($S_{c}$ and $S_{em}$ are the time integrals of $L_{c}$ and $L_{em}$
respectively).

The trace in eqn.~(\ref{treqn}) is evaluated perturbatively
\begin{eqnarray}
{\rm Tr}\,\ln G^{-1} & = &  {\rm Tr}\,\ln \left(\,G^{-1}_{0} + \delta G^{-1}\,
                         \right) \nonumber \\
               & = &  {\rm Tr}\,\ln G^{-1}_{0} +{\rm Tr}\, G_{0}\delta G^{-1}
                       -\frac{1}{2}{\rm Tr}\, G_{0}\delta G^{-1}G_{0}
                        \delta G^{-1}
                        + \cdots \hspace{0.1in} .
\label{perturb}
\end{eqnarray}
For the purpose of finding the hydrodynamic action, it is only necessary to
work out the expansion in eqn.~(\ref{perturb}) to second order in space/time
derivatives of the gap phase. Thus
\begin{eqnarray}
S & = & S_{0}+S_{1}+S_{2}+\cdots \nonumber \\
  & = & S_{0}+S_{hyd}+\cdots  \hspace{0.1in} .
\end{eqnarray}
Eckern et.\ al.\ \cite{ambe} have determined $S_{0}$ and $S_{2}$. $S_{0}$ is
the action for the bulk degrees of freedom. It is not of immediate interest to
us, and so will not be discussed further. $S_{2}$ is found to be
\begin{equation}
S_{2}= \int\, dt\,d^{3}x\, \left[\, \frac{m}{2}\rho_{s}{\bf v}_{s}^{2} +
        N(0)\tilde{A}_{0}^{2}\,\right] \hspace{0.1in} ,
\end{equation}
where $m$ has been re-instated;
$\rho_{s}$ is the density of superconducting electrons at
$T=0$; and $N(0)$ is the electron density of states at the Fermi surface.

It is our purpose to show that $S_{1}$ is the gauge invariant Wess-Zumino
term found in Section~\ref{sec2c}, where
\begin{equation}
S_{1}=S_{t1}+S_{s1}= i\, {\rm Tr}\, G_{0}\left( \delta G^{-1}_{t1} +
        \delta G^{-1}_{s1}\right) \hspace{0.1in} .
\end{equation}
We will evaluate $S_{t1}$ below. A similar calculation (which will not be
reproduced here) shows that $S_{s1}$ vanishes. Thus we focus on
\begin{eqnarray}
S_{t1} & = & {\rm Tr}\, G_{0}\delta G^{-1}_{t1} \nonumber \\
       & = & i\int\, d^{3}x\, d^{3}y\, {\rm tr}\,\langle\, x| G_{0}|y\,\rangle
              \langle\, y|\delta G^{-1}_{t1}|x\,\rangle \nonumber \\
       & = & i\int\, d^{3}x\, d^{3}y\, {\rm tr}\left\{ \, G_{0}(x-y)
              \left[ \tilde{A}_{0}(y)\sigma_{3}\,\delta^{3}({\bf x}-{\bf y})
               \delta(t_{x}-t_{y})\right]\,\right\} \hspace{0.1in} ,
\label{pert1}
\end{eqnarray}
and ${\rm tr}$ is a sum over Nambu indices only. We are interested in
evaluating eqn.~(\ref{pert1}) to lowest order in gradients of the gap phase.
As $\tilde{A}_{0}$ is already first order in gradients, we can write
\begin{equation}
\tilde{A}_{0}({\bf y}) = \tilde{A}_{0}({\bf R} -
  \mbox{\boldmath $\rho$}/2) \approx \tilde{A}_{0}
   ({\bf R}) \hspace{0.1in} ,
\end{equation}
where ${\bf R}=({\bf x} + {\bf y})/2$; and $\mbox{\boldmath $\rho$} =
{\bf x} - {\bf y}$. Also, $G_{0}(x-y)$ can be written as
\begin{equation}
G_{0}(x-y)= \int\, \frac{d^{4}k}{(2\pi)^{4}}\,\frac{e^{ik(x-y)}}
             {k_{0}- ( ({\bf k}^{2}/2) - E_{f})\sigma_{3} - \Delta_{0}
               \sigma_{1} } \hspace{0.1in} .
\end{equation}
Due to the vortex, $\Delta_{0}$ varies with position near the vortex core. In
the local limit, $\Delta_{0}$ is constant except on a set of measure zero
(viz.\ on the vortex core). We will approximate $\Delta_{0}$ by its constant
value away from the vortex core. Putting all this together gives
\begin{eqnarray}
S_{t1} & = & \int\, d^{3}R\, d^{3}\rho\, i\, {\rm tr}\int\, \frac{d^{4}k}
              {(2\pi)^{4}}\, {\frac{e^{ik(x-y)}}{k_{0} - (({\bf k}^{2}/2)-
               E_{f})\sigma_{3} - \Delta_{0}\sigma_{1}}}
               \tilde{A}_{0}({\bf R})\,\sigma_{3}
                \int\,\frac{d^{4}k^{\prime}}{(2\pi)^{4}}\, e^{-ik^{\prime}
                 (x-y)} \nonumber \\
       & = & -\int\, d^{3}R\, \tilde{A}_{0}({\bf R})\,\int\,\frac{d^{4}k}
              {(2\pi)^{4}}\,{\rm tr}\left[\, -i G_{0}({\bf k})\sigma_{3}\,
               \right]
               \nonumber \\
       & = & \int\, d^{3}R\, \rho_{s}\left[\, \frac{\hbar}{2}\partial_{t}
              \theta - e A_{0}\,\right] \hspace{0.1in} ,
\end{eqnarray}
where $\hbar$ has been re-instated. Thus, $S_{t1}$ is the gauge invariant
Wess-Zumino term found in Section~\ref{sec2c}. Since $S_{s1}=0$, we have
$S_{1}=S_{t1}$ and
\begin{eqnarray}
S_{hyd} & = & \int\, d^{3}x\,dt\, \rho_{s}\left[\, \frac{\hbar}{2}\partial_{t}
               \theta - eA_{0}\,\right] \nonumber \\
        &   & {} \hspace{0.25in} + \int\, d^{3}x\, dt\, \left[ \, \frac{m}{2}
                \rho_{s}{\bf v}_{s}^{2} + N(0)\left( \, \frac{\hbar}{2}
                 \partial_{t}\theta -eA_{0}\,\right)^{2}\,\right]
        \hspace{0.1in} .
\end{eqnarray}

We stress that the calculation of the Wess-Zumino term given in this Appendix
is {\em independent\/} of the Berry phase calculation given in
Section~\ref{sec2c}
and thus acts to confirm its result, which is that $S_{hyd}$ contains a
gauge invariant term first order in time derivatives of the gap phase. The
occurrence of this term in $S$ appears to have been overlooked in
Ref.~\onlinecite{ambe}.


\begin{references}
\bibitem{bs} J. Bardeen and M. J. Stephen, Phys.\ Rev.\ {\bf 140}, A1197
             (1965).
\bibitem{nv} P. Nozi\`{e}res and W. F. Vinen, Phil. Mag. {\bf 14}, 667 (1966).
\bibitem{bat} G. K. Batchelor, ``An Introduction to Fluid Dynamics'',
              pp.\ 404-407 (Cambridge University Press, New York 1967).
\bibitem{ks} Y. B. Kim and M. J. Stephen, in ``Superconductivity'',
             R. D. Parks, ed.\ , (Marcel Decker, New York 1969);
             W. F. Vinen, in ``Superconductivity'', R. D. Parks, ed.\ ,
             (Marcel Decker, New York 1969).
\bibitem{ath} P. Ao and D. J. Thouless, Phys.\ Rev.\ Lett.\ {\bf 70}, 2158
              (1993).
\bibitem{deg} P. G. deGennes, ``Superconductivity of Metals and Alloys'',
              chap.\ 5 (Addison-Wesley Publishing Company, Inc. New York
              1989).
\bibitem{us} W. E. Goff, F. Gaitan and M. Stone, Phys.\ Lett.\ A {\bf 136},
             433 (1989).
\bibitem{nam} Y. Nambu, Phys.\ Rev.\ {\bf 117}, 648 (1960).
\bibitem{bar} J. Bardeen, R. Kummel, A. E. Jacobs and L. Tewordt, Phys.\
              Rev.\ {\bf 187}, 556 (1969).
\bibitem{ber} M. V. Berry, Proc.\ R. Soc.\ London A {\bf 392}, 45 (1984).
\bibitem{ahar} Y. Aharanov and J. Anandan, Phys.\ Rev.\ Lett.\ {\bf 58},
               1593 (1987).
\bibitem{car} C. Caroli, P. G. deGennes and J. Matricon, Phys.\ Lett.\ {\bf 9},
              307 (1964).
\bibitem{schr} J. R. Schrieffer, ``Theory of Superconductivity'', p.\ 59
               (Addison-Wesley Publishing Company, Inc.\ New York 1964).
\bibitem{pwa} P. W. Anderson, Rev.\ Mod.\ Phys.\ {\bf 38}, 298 (1966).
\bibitem{f+h} A. L. Fetter and P. C. Hohenberg, in ``Superconductivity'',
              R. D. Parks, ed.\ (Marcel Decker, New York 1969).
\bibitem{hs} R. L. Stratonovitch, Dokl.\ Akad.\ Nauk.\ S. S. R. {\bf 115},
             1907 (1957) [Sov.\ Phys.\ Dokl.\ {\bf 2}, 416 (1958)];
             J. Hubbard, Phys.\ Rev.\ Lett.\ {\bf 3}, 77 (1959).
\bibitem{bard2} J. Bardeen, Phys.\ Rev.\ Lett.\ {\bf 13}, 747 (1964).
\bibitem{ambe} U. Eckern, G. Schon and V. Ambegaokar, Phys.\ Rev.\ B {\bf 30},
               6419 (1984).
\bibitem{wz} J. Wess and B. Zumino, Phys. Lett. B {\bf 37}, 95 (1971);
             A. Shapere and F. Wilczek, ``Geometric Phases in Physics'',
             chap.\ 7 (World Scientific Publishing Co. Singapore 1989).
\bibitem{eik} A. Garg, V. P. Nair and M. Stone, Ann.\ Phys.\ {\bf 173}, 149
              (1987); F. Gaitan, Ann.\ Phys.\ (to be published); F. Gaitan,
              Phys.\ Lett.\ A {\bf 151}, 551 (1991).
\bibitem{fuj} K. Fujikawa, Phys.\ Rev.\ D {\bf 21}, 2848 (1980).
\bibitem{gw} J. Goldstone and F. Wilczek, Phys.\ Rev.\ Lett.\ {\bf 47}, 986
             (1981).
\bibitem{fg2} A similar calculation was done in the last paper cited in
              Ref.~\onlinecite{eik}.
\bibitem{chrv} C. G. Callan and J. A. Harvey, Nuc.\ Phys.\ B {\bf 250},
               427 (1985).
\bibitem{hat} M. Hatsuda, S. Yahikozawa, P. Ao and D. J. Thouless,
              Phys.\ Rev.\ B {\bf 49}, 15870 (1994) propose a Ginzburg-Landau
              action in which a topological term is introduced by hand. This
              contrasts with the approach taken in this paper in which we
              derive the Wess-Zumino term from a microscopic starting point.
\bibitem{lobb} C. J. Lobb, invited talk presented at 1994 APS March Meeting
               (Pittsburgh, Pa.\ ).
\bibitem{exper} Y. Iye, S. Nakamura and T. Tamegai, Physica C {\bf 159},
                616 (1989); S. N. Artemenko, I. G. Gorlova and Yu.\ I.
                Latyshev, Phys.\ Lett.\ A {\bf 138}, 428 (1989); S. J. Hagen,
                C. J. Lobb, R. L. Greene, M. G. Forrester and J. H. Kang,
                Phys.\ Rev.\ B {\bf 41}, 11630 (1990); S. J. Hagen, C. J. Lobb,
                R. L. Greene and M. Eddy, Phys.\ Rev.\ B {\bf 43}, 6246 (1991);
                T. R. Chien, T. W. Jing, N. P. Ong, and Z. Z. Wang, Phys.\
                Rev.\ Lett.\ {\bf 66}, 3075 (1991); J. Luo, T. P. Orlando,
                J. M. Graybeal, X. D. Wu and R. Muenchausen, Phys.\ Rev.\
                Lett.\ {\bf 68}, 690 (1992).
\bibitem{vol} G. E. Volovik, Zh.\ Eksp.\ Teor.\ Fiz.\ {\bf 104}, 3070 (1993)
              [Sov.\ Phys.\ JETP {\bf 77}, 435 (1993)].
\bibitem{ao2} P. Ao, Q. Niu and D. J. Thouless, Physica B {\bf 194-196}, 1453
              (1994).
\bibitem{mes} A. Messiah, ``Quantum Mechanics'', Vol.\ II (North-Holland
              Publishing Company. Amsterdam 1961).
\end{references}
\end{document}